\documentclass[]{interact}

\usepackage{epstopdf}
\usepackage[caption=false]{subfig}

\usepackage[numbers,sort&compress]{natbib}
\bibpunct[, ]{[}{]}{,}{n}{,}{,}
\renewcommand\bibfont{\fontsize{10}{12}\selectfont}
\makeatletter
\def\NAT@def@citea{\def\@citea{\NAT@separator}}
\makeatother

\theoremstyle{plain}
\newtheorem{theorem}{Theorem}[section]

\theoremstyle{definition}

\theoremstyle{remark}
\newtheorem{remark}{Remark}

\usepackage{algorithm}
\usepackage{algorithmic} 

\usepackage{my}
\usepackage{color}
\usepackage{booktabs}

\usepackage{ascmac}
\allowdisplaybreaks
 
\newcommand{\rfig}[1]{Fig.~\ref{#1}} 
\def\qed{\hfill $\blacksquare$}

\newtheorem{problem}[theorem]{Problem}
\newtheorem{assumption}[theorem]{Assumption}

\usepackage{hyperref}

\begin{document}

\title{Two-step reinforcement learning for model-free redesign of nonlinear optimal regulator}

\author{
\name{Mei Minami\textsuperscript{a}, 
Yuka Masumoto\textsuperscript{a}, 
Yoshihiro Okawa\textsuperscript{b},
Tomotake Sasaki\textsuperscript{b} and
Yutaka Hori\textsuperscript{a}\thanks{Mail : yhori@appi.keio.ac.jp}}
\affil{\textsuperscript{a}
Department of Applied Physics and Physico-Informatics, Keio University, Japan;
\textsuperscript{b}Artificial Intelligence Laboratory, Fujitsu Limited, Japan}}

\maketitle

\begin{abstract}
In many practical control applications, the performance level of a closed-loop system degrades over time due to the change of plant characteristics. 
Thus, there is a strong need for redesigning a controller without going through the system modeling process, which is often difficult for closed-loop systems.
Reinforcement learning (RL) is one of the promising approaches that enable  model-free redesign of optimal controllers for nonlinear dynamical systems based only on the measurement of the closed-loop system. However, the learning process of RL usually requires a considerable number of trial-and-error experiments using the poorly controlled system that may accumulate wear on the plant. 
To overcome this limitation, we propose a model-free two-step design approach that improves the transient learning performance of RL in an optimal regulator redesign problem for unknown nonlinear systems. Specifically, we first design a linear control law that attains some degree of control performance in a model-free manner, and then, train the nonlinear optimal control law with online RL by using the designed linear control law in parallel. We introduce an offline RL algorithm for the design of the linear control law and theoretically guarantee its convergence to the LQR controller under mild assumptions. Numerical simulations show that the proposed approach improves the transient learning performance and efficiency in hyperparameter tuning of RL.
\end{abstract}


\section{INTRODUCTION}
In many practical control applications, a reliable mathematical model of the controlled plant is not available since system modeling is often difficult.
A common example of such situations arises when the plant is already in use in a closed-loop system for daily operations and is hard to suspend despite a degraded performance level due to the change of plant characteristics.
This leads to a strong need for redesigning or tuning the controller based only on the measurement of the closed-loop system in a model-free manner. 
Until now, model-free controller design methods have been developed using various approaches for linear and nonlinear systems. Examples include 
data-driven control (DDC)\cite{Hou2013},  iterative learning control (ILC), model-free adaptive control (MFAC)\cite{Hou2019}, iterative feedback tuning (IFT)\cite{Hjalmarsson1998}, virtual reference feedback tuning (VRFT)\cite{Campi2002}, and fictitious reference iterative tuning (FRIT)\cite{Kaneko2013data}.
Among them, reinforcement learning (RL) \cite{sutton2018reinforcement,Lewis2012_RL_and_FC,optimal_and_autonomous_control_using_RL} has been actively studied as a versatile approach to tackle optimal control problems for a large class of nonlinear systems in recent years.
An attracting feature of RL-based controller design is that the optimal control law is explored in a fully autonomous fashion through a trial-and-error process \cite{bradtke1994adaptive,doya2000reinforcement,murray2002adaptive,Doya2002Multiple,Al-Tamimi2008,vamvoudakis2010online,sprangers2015reinforcement}.

One of the major issues of the RL-based design is that the system usually needs to undergo many trial-and-error experiments using the poorly designed control law during the learning process, which may accumulate wear on the plant and potentially shortens the lifetime of the system. Thus, it is desirable to develop a learning method that (i) maintains the performance level of the closed-loop system to some extent during the learning process, and (ii) reduces the number of necessary trials.
Several studies have tackled these problems by utilizing LQR controllers designed based on the mathematical model of the controlled object.
For example, literature \cite{Amherst2000} proposed to switch a local LQR controller and an RL controller depending on whether the current state of the system is inside the estimated controllable set or not,
while literature \cite{okawa2019control} proposed to use both the  RL and LQR controllers at the same time in parallel to improve its transient learning performance. In addition, literature \cite{Samuele2021RL} also used RL to learn a nonlinear controller operated with the LQR controller, in such a way that the derivative of the nonlinear controller with respect to the state becomes zero at the origin to ensure the local stability. 
Futhermore, methods that combine RL and model predictive control (MPC) were  proposed in literature \cite{zanon2020safe,xie2020model} to tackle the safety, stability, and trial amount (sample efficiency) issues in the exploration phase of RL.

However, these methods require some form of the plant model, and thus, their application to the closed-loop controller {\it redesign} problem is limited.
This motivates us to further develop a model-free approach for assisting the learning process of RL.

In this paper, we propose a completely model-free approach to design an optimal control law, especially a nonlinear optimal regulator, for nonlinear systems while improving the transient learning performance and efficiency of RL. 
Specifically, we consider a situation where a closed-loop system with a stabilizing linear controller has an unsatisfactory performance level and requires redesign of the control law without knowing  the mathematical model of the plant.
The proposed approach consists of two steps. In the first step, we measure the input-output response of the existing closed-loop system and design a quasi-optimal linear quadratic regulator (LQR) in an offline and model-free manner, which achieves a certain degree of performance to assist the learning process of RL. 
Then, in the second step, we use an online RL method to design a nonlinear control law that is connected, in parallel, to the pre-designed linear control law. As a result of this two-step design approach, the designed control law achieves a performance that cannot be realized by the linear control law alone.
The proposed approach has two main advantages for its practical use: (i) it can be applied even if the plant model is unavailable, and (ii) it can reduce wear on the actual plant during the learning process of RL because the performance level can be improved by the quasi-optimal LQR controller, and the efficiency in hyperparameter tuning is improved.

The organization of this paper is as follows. In Section \ref{sec:problem_formulation}, we address the problem formulation. In Section \ref{sec:controlaw}, we describe the proposed approach. In Section \ref{sec:step1}, we introduce the algorithm for Step~1 of the proposed approach.
Then, in Section \ref{sec:example}, the effectiveness of the proposed approach is verified by numerical simulations of an inverted pendulum with input saturation as the plant. Section \ref{sec:robustness} illustrates the advantage of the proposed approach that hyperparameters can be efficiently tuned. Finally, in Section \ref{sec:conclusion},  we give concluding remarks of this paper.

\section{PROBLEM FORMULATION}
\label{sec:problem_formulation}
In this section, we describe the problem formulation.
We consider a discrete-time nonlinear system described by 
\begin{align}
    \bm{x}_{k+1}=\bm{f}(\bm{x}_{k},\bm{u}_{k}), 
\label{Mnonlinear}
\end{align}
where $\bm{x}_{k} \in \mathbb{R}^n$ and $\bm{u}_{k} \in \mathbb{R}^m$ are the state and input at time $k$, and $\bm{f} : \mathbb{R}^n \times \mathbb{R}^m \rightarrow \mathbb{R}^n$ is a nonlinear function. The equilibrium point of interest is at the origin, and $\bm{f}$ is smooth at that point. The state $\bm{x}_{k}$ can be measured directly.

We define $A \in \mathbb{R}^{n \times n}$ and $B \in \mathbb{R}^{n \times m}$ by
\begin{align}
A := \frac{\partial \bm{f}}{\partial \bm{x}} \biggm\vert_{\bm{x} = \bm{0}, \bm{u} = \bm{0}} ,  B := \frac{\partial \bm{f}}{\partial \bm{u}} \biggm\vert_{\bm{x} = \bm{0}, \bm{u} = \bm{0}}.
\end{align}
Using $A$ and $B$, the linear approximation of the system dynamics near the origin is given by
\begin{align}
    \bm{x}_{k+1}=A\bm{x}_k+B\bm{u}_{k}.
    \label{linear_system}
\end{align}

In what follows, we consider the case where the nonlinear plant (\ref{Mnonlinear}) is already in operation with a locally stabilizing linear control law $K^{\mathrm{init}}$, but the performance of the closed-loop system has room for improvement, and the plant model is unknown. 
Then, our goal is to redesign the control law (policy) for better performance, but without explicitly identifying the nonlinear function $\bm{f}$ and its associated Jacobian matrices $A$ and $B$. 

This situation can be more formally stated as follows.
\begin{assumption}
A linear control law $K^{\mathrm{init}}\in\mathbb{R}^{m \times n}$ such that $A + BK^{\mathrm{init}}$ is Schur stable is given. \label{supp1}
\end{assumption}

With this assumption, we consider the following nonlinear optimal regulator design problem. 

\begin{problem}\label{problem}
Consider the nonlinear system (\ref{Mnonlinear}). Suppose Assumption \ref{supp1} holds, and the plant model of (\ref{Mnonlinear}) is unknown.
Let
\begin{align}
    J=\sum^{\infty}_{k=0} (\bm{x}_{k}^\top Q \bm{x}_{k} + \bm{u}_{k}^\top R \bm{u}_{k}), \label{costfunc}
\end{align}
where $Q \in \mathbb{R}^{n \times n}$ and $R \in \mathbb{R}^{m \times m}$ are given positive semi-definite and positive definite symmetric matrices, respectively.
Design a control law that minimizes the cost function \eqref{costfunc}. 
\end{problem}

Note that the solution of Problem \ref{problem} (\textit{i.e.,} optimal regulator) is nonlinear, because the system dynamics is so. 
\section{PROPOSED APPROACH: MODEL-FREE TWO-STEP DESIGN OF CONTROL LAW}
\label{sec:controlaw}
\begin{figure}[tb]
  \centering
  \includegraphics[width=10cm]{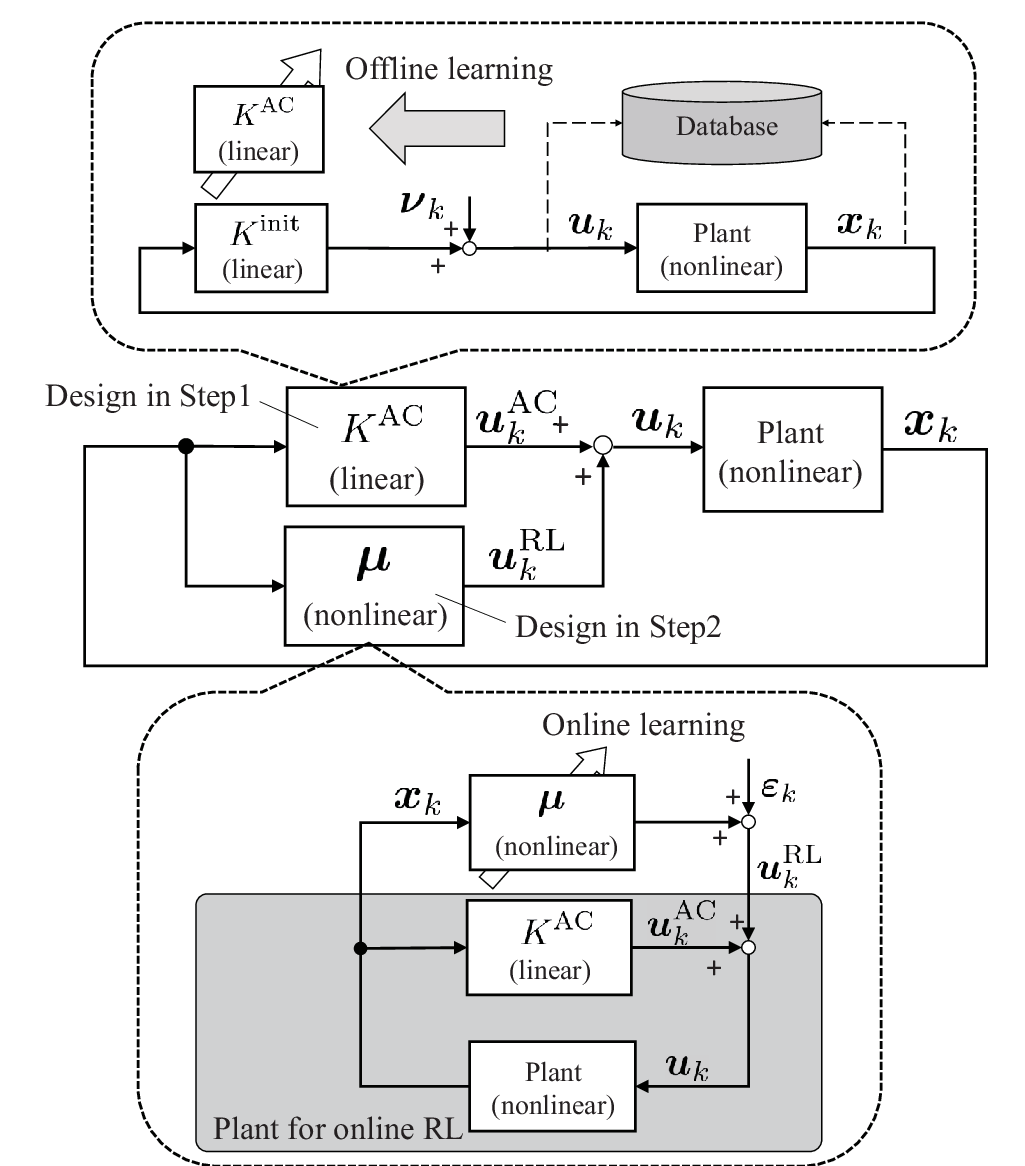}
  \caption{Structure of the proposed control law and the model-free two-step design approach. In Step 1, the linear auxiliary control law $K^\mathrm{AC}$ is designed by offline RL, which contributes to reducing wear on the plant in the learning process of  online RL. In Step 2, we design the nonlinear control law ${\bm \mu}$ by online RL.}
  \label{fig:model_free_two_step_design}
\end{figure}
We propose a model-free two-step approach to design the optimal control law.
The structure of the control law we use in the proposed approach is shown in \rfig{fig:model_free_two_step_design}, where $K^\mathrm{AC}$ is a linear auxiliary control (AC) law that assists the learning process of a nonlinear control law $\bm{\mu}: \mathbb{R}^{n} \times \mathcal{W} \to \mathbb{R}^{m}$ with an adjustable parameter $W \in \mathcal{W}$ ($\mathcal{W}$: a set of parameters). The proposed approach is to design the parallel control laws $K^\mathrm{AC}$ and $\bm{\mu}$ by  the following two-step procedure.
\begin{itembox}{Two-step design procedure}
\textbf{Step 1:}~Design the auxiliary control law $K^\mathrm{AC} \in \mathbb{R}^{m \times n}$ for the linear quadratic regulator (LQR) problem using an offline RL method. \\
\textbf{Step 2:}~Design the nonlinear control law $\bm{\mu}$ (adjust the parameter $W \in \mathcal{W}$) by an online RL method, where we regard the closed-loop system consisting of the plant and the auxiliary control law $K^\mathrm{AC}$ as a single environment (plant for online RL shown as a gray box in \rfig{fig:model_free_two_step_design}).
\end{itembox}

The linear auxiliary control law $K^\mathrm{AC}$ designed in Step 1 makes the cost lower than the initial control law $K^\mathrm{init}$ while it is only quasi-optimal due to the nonlinearity of the plant. This new linear  control law contributes to reducing wear on the plant in the learning process of the online RL in Step 2. 
This is because (i) the auxiliary control law improves the transient learning performance of the online RL by attaining a certain level of performance at the early stage of learning, and moreover, (ii) it facilitates the hyperparameter tuning of online RL by avoiding the application of large inputs from RL. 
It should be noted that the damage to the plant by Step 1 can be considered to be minimal since the linear control law designed in Step 1 is obtained by an offline algorithm using a single set of input-output data collected with the pre-existing control law $K^{\mathrm{init}}$.

Then, in Step 2, the nonlinear control law ${\bm \mu}$ is designed by the trial-and-error process assisted by the auxiliary control law $K^{\mathrm{AC}}$. Specifically, the control input $\bm{u}_{k}$ is generated by
\begin{align}
\bm{u}_{k} = \bm{u}^{\mathrm{AC}}_{k} + \bm{u}^{\mathrm{RL}}_{k}, \label{crlmbc7}
\end{align}
where $\bm{u}^{\mathrm{AC}}_{k} = K^{\mathrm{AC}} \bm{x}_{k}$, and $\bm{u}^{\mathrm{RL}}_{k}$ is the control input obtained by an online RL method based on $\bm{\mu}$. 
The nonlinear control law ${\bm \mu}$ enables to further lower the value of the cost function and attains the performance level that the linear controllers cannot achieve. 
Although this improvement requires additional cost for the learning of the control law $\bm{\mu}$, the effect of the improvement is significant in the long run since the designed control law is repeatedly used in the actual operation of the plant.

The underlying spirit of the two-step design approach is similar to the residual learning \cite{he2016deep, silver2018residual}, which is a widely used approach in machine learning, but the proposed approach is specifically tailored for redesigning an optimal regulator for dynamical systems. 
In the two-step design procedure, the offline and the online RL methods for the design of linear and nonlinear control laws can be freely chosen by users. 
Nevertheless, in the next section, we show a specific instance of the design method of the linear auxiliary control law in Step 1. Then, in Section \ref{sec:example}, we demonstrate how the linear auxiliary control law can be combined with an online RL method to train the nonlinear controller at a relatively small cost.

\section{OFFLINE REINFORCEMENT LEARNING FOR DESIGNINIG LINEAR AUXILIARY CONTROL LAW}
\label{sec:step1}
The algorithm for offline RL in Step 1 is given in Algorithm \ref{alg1}.
This algorithm is derived based on the Hewer's iterative method for discrete-time algebraic Riccati equation~\cite{Difference_Riccati_equation_convergence}. The algorithm executes the Hewer's method by using collected input-output data instead of the model of the plant, \textit{i.e.,} the matrices $A$ and $B$ in \eqref{linear_system}, and finds the LQR controller for the linearized system. 
First, as shown in the upper part of \rfig{fig:model_free_two_step_design}, we apply an exploration term $\bm{\nu}_{k}$  to the closed-loop system locally stabilized by the control law  $K^{\mathrm{init}}$, and record the input $\bm{u}_{k}$ and output $\bm{x}_{k}$. 
We denote the collected time series data by $\{\rvu_k \}_{k = k_{\mathrm{s}}}^{k_{\mathrm{s}} + l -1}$ and $\{\rvx_k \}_{k = k_{\mathrm{s}}}^{k_{\mathrm{s}} + l }$ where the non-negative integer $k_{\mathrm{s}}$ is the arbitrarily determined start time of data collection, and $k_{\mathrm{s}}+l$ is the final time of input-output data.
Then, the auxiliary control law $K^\mathrm{AC}$ which assists the learning process of the online RL is obtained by the iteration of (I) and (II) in Algorithm~\ref{alg1}, where 
 $\bm{h}^{j} \in\mathbb{R}^l$ and $F^{j} \in \mathbb{R}^{l\times (n^2 + nm + m^2)}$ are defined by 
\begin{align}
\bm{h}^j 
:=&[h^j_0,h^j_1,\ldots ,h^j_{l-1}]^\top
\label{phi},\\
F^{j}:=&\begin{bmatrix}
F^{j}_{(xx)0} & F^{j}_{(xu)0} & F^{j}_{(uu)0} \\
F^{j}_{(xx)1} & F^{j}_{(xu)1} & F^{j}_{(uu)1} \\
\vdots & \vdots & \vdots \\
F^{j}_{(xx)l-1} & F^{j}_{(xu)l-1} & F^{j}_{(uu)l-1} \\
\end{bmatrix} 
\label{psi}
\end{align}
with the entries
\begin{align}
    h^j_i :=& \rvx_{k_{\mathrm{s}}+i}^\top Q \rvx_{k_{\mathrm{s}}+i}+ \rvx_{k_{\mathrm{s}}+i}^\top (K^j)^\top R K^j \rvx_{k_{\mathrm{s}}+i},  \\
F^{j}_{(xx)i}:=&\rvx_{k_{\mathrm{s}}+i}^\top \otimes \rvx_{k_{\mathrm{s}}+i}^\top - \rvx_{k_{\mathrm{s}}+i+1}^\top \otimes \rvx_{k_{\mathrm{s}}+i+1}^\top, \label{F^{j}_{(xx)i}} \\
F^{j}_{(xu)i}:=&2(\rvx_{k_{\mathrm{s}}+i}^\top \otimes (\rvu_{k_{\mathrm{s}}+i}-K^j \rvx_{k_{\mathrm{s}}+i})^\top),  \\
F^{j}_{(uu)i}:=&(\rvu_{k_{\mathrm{s}}+i}+K^j \rvx_{k_{\mathrm{s}}+i} )^\top \otimes ( \rvu_{k_{\mathrm{s}}+i}-K^j\rvx_{k_{\mathrm{s}}+i})^\top. \label{F^{j}_{(uu)i}} 
\end{align}
The symbols $\otimes$ in \eqref{F^{j}_{(xx)i}} to \eqref{F^{j}_{(uu)i}} and $\mathrm{vec}$ in \eqref{algorithm_AC} in Algorithm \ref{alg1} are the Kronecker product and 
 the vec operator, respectively~\cite[\S 4]{hj1994topics}.
\begin{algorithm}[t]                      
\caption{Offline RL Algorithm for Discrete-Time LQR Problem.}    
\label{alg1}   
\textbf{Data Collection.} Apply $ \bm{u}_{k}=K^{\mathrm{init}} \bm{x}_{k} + \bm{\nu}_k $ to the
plant to collect data for $ k = k_{\mathrm{s}},k_{\mathrm{s}}+1,k_{\mathrm{s}}+2,\ldots ,k_{\mathrm{s}}+l$, where $\bm{\nu}_k \in \mathbb{R}^m $ is an exploration term.\\
\textbf{Initialization.} Set the iteration number $j=0$ and the linear control law $K^0=K^{\mathrm{init}}$.\\
\textbf{Policy Evaluation and Improvement.} Based on the collected data, perform the following iterations for $j = 0,1, \dots$.

$\rm(\,I\,)$ Calculate $G^{j}_1 \in \sR^{n \times n}, G^{j}_2 \in \sR^{m \times n},G^{j}_3 \in \sR^{m \times m}$ that give a least square solution of   
the following equation, where $\bm{h}^{j}$ is \eqref{phi} and $F^{j}$ is \eqref{psi}: 
\begin{align}
  F^{j}[\textrm{vec}(G^{j}_1)^\top ,\textrm{vec}(G^{j}_2)^\top ,\textrm{vec}(G^{j}_3)^\top]^\top =  \bm{h}^j. \label{algorithm_AC}
\end{align}
$\rm(I\hspace{-.01em}I)$ Update the linear control law by the following equation: 
\begin{align}
K^{j+1}=-(G^{j}_3 + R)^{-1}G^{j}_2. \label{al1_K}
\end{align}
\textbf{Repetition and Termination.} Repeat the policy evaluation and improvement with $j \leftarrow j+1$ until
\begin{align}
\|K^{j+1}-K^{j}\| \leq \epsilon \label{termination_condition}
\end{align}
is satisfied for a small positive scalar $\epsilon$. 
\end{algorithm}

As shown in Theorem~\ref{th:1} below, the convergence of $K^j$ in Algorithm~\ref{alg1} to the LQR solution is guaranteed when the column rank of $F^{j}$ is  greater than or equal to $n^2 + nm + m^2 $ for all $j=0,1,\ldots$, in which case the solution of the linear equation \eqref{algorithm_AC} is uniquely determined. 
This condition 
can be satisfied by choosing $l$ such that  
\begin{align}
    l \geq n^2 + nm + m^2
    \label{condition_of_l}
\end{align}
and adding an appropriate exploration term that sufficiently excites the system.

\begin{theorem}\label{th:1}
Consider a linear system whose state equation is \eqref{linear_system}.
Suppose Assumption \ref{supp1}
holds, and $\mathrm{rank}(F^{j}) = n^2 + m^2 + nm$ for all $j=0,1,\ldots$.
Then, $K^j$ updated by Algorithm~\ref{alg1} converges to the LQR solution $K^\star$ of the discrete-time linear system \eqref{linear_system}
as $j \rightarrow \infty $. Furthermore, the rate of convergence is quadratic in the neighborhood of the solution $K^\star$.
\end{theorem}
\begin{proof}
We begin with the case $j = 0$. Since $A + BK^{\mathrm{init}}$ is Schur stable (Assumption \ref{supp1}),  there exists a positive definite symmetric matrix $P^{0} \in \mathbb{R}^{ n \times n } $  satisfying
\begin{align}
\vx_{k}^\top P^{0} \vx_{k} = \sum^{\infty}_{i=k}(\vx_{i}^\top Q \vx_{i} +  (K^{\mathrm{init}} \vx_i)^\top R K^{\mathrm{init}} \vx_i). \label{dfn_P0}
\end{align}
About this equation, the following relation holds: 
\begin{align}
\mathrm{RHS~of~} \eqref{dfn_P0} & = \vx_{k}^\top Q \vx_{k} + (K^{\mathrm{init}} \vx_k)^\top R K^{\mathrm{init}} \vx_k \nonumber 
 + \sum^{\infty}_{i=k+1}(\vx_{i}^\top Q \vx_{i} +  (K^{\mathrm{init}} \vx_i)^\top R K^{\mathrm{init}} \vx_i) \nonumber \\ 
& = \vx_{k}^\top ( Q + (K^{\mathrm{init}})^\top R K^{\mathrm{init}} ) \vx_{k}
 + \vx_{k+1}^\top P^{0} \vx_{k+1}.
\notag
\end{align}
This leads to the following Bellman equation: 
\begin{align}
\vx_{k}^\top P^{0} \vx_{k} & = \vx_{k}^\top ( Q + (K^{\mathrm{init}})^\top R K^{\mathrm{init}} ) \vx_{k} + \vx_{k+1}^\top P^{0} \vx_{k+1}. \label{Bellman}
\end{align}
Under the control law $ K^{\mathrm{init}} $, the following equation holds:
\begin{align}
\vx_{k+1} = A \vx_{k} + B K^{\mathrm{init}} \vx_{k}. \label{StateEquationFB}
\end{align}
This can be transformed as below with an arbitrary $\vu_k$: 
\begin{align}
\vx_{k+1} & = A \vx_{k} + B K^{\mathrm{init}}\vx_{k} + B\vu_k - B\vu_k  \nonumber \\
& = (A \vx_{k} + B \vu_k) - B (\vu_k - K^{\mathrm{init}} \vx_{k}).
\end{align}
Substituting this equation and the collected data $\rvu_{k_{\mathrm{s}}}$ and  $\rvx_{k_{\mathrm{s}}}$ into the Bellman equation~\eqref{Bellman}, 
we have
\begin{align}
\rvx_{k_{\mathrm{s}}}^\top P^{0} \rvx_{k_{\mathrm{s}}} 
& = \rvx_{k_{\mathrm{s}}}^\top ( Q + (K^{\mathrm{init}})^\top R K^{\mathrm{init}} ) \rvx_{k_{\mathrm{s}}} \nonumber \\
& {} + \rvx_{k_{\mathrm{s}} +1}^\top P^{0} \rvx_{k_{\mathrm{s}} +1}  \nonumber \\
& {} - (A \rvx_{k_{\mathrm{s}}} + B \rvu_{k_{\mathrm{s}}})^\top P^{0} B (\rvu_{k_{\mathrm{s}}} - K^{\mathrm{init}} \rvx_{k_{\mathrm{s}}})  \nonumber \\
& {} - (\rvu_{k_{\mathrm{s}}} - K^{\mathrm{init}} \rvx_{k_{\mathrm{s}}})^\top B^\top P^{0} (A \rvx_{k_{\mathrm{s}}} + B \rvu_{k_{\mathrm{s}}}) \nonumber \\
& {} + (\rvu_{k_{\mathrm{s}}} - K^{\mathrm{init}} \rvx_{k_{\mathrm{s}}})^\top B^\top P^{0} B (\rvu_{k_{\mathrm{s}}} - K^{\mathrm{init}} \rvx_{k_{\mathrm{s}}}) \nonumber \\
& = \rvx_{k_{\mathrm{s}}}^\top ( Q + (K^{\mathrm{init}})^\top R K^{\mathrm{init}} ) \rvx_{k_{\mathrm{s}}} \nonumber \\ 
& {} + \rvx_{k_{\mathrm{s}} +1}^\top P^{0} \rvx_{k_{\mathrm{s}} +1}  \nonumber \\
& {} - 2 (\rvu_{k_{\mathrm{s}}} - K^{\mathrm{init}} \rvx_{k_{\mathrm{s}}})^\top B^\top P^{0} A \rvx_{k_{\mathrm{s}}}   \nonumber \\
& {} - (\rvu_{k_{\mathrm{s}}} - K^{\mathrm{init}} \rvx_{k_{\mathrm{s}}})^\top B^\top P^{0} B (\rvu_{k_{\mathrm{s}}} + K^{\mathrm{init}} \rvx_{k_{\mathrm{s}}}). \label{Bellman2.5}
\end{align}
\noindent
We partly used $A \rvx_{k_{\mathrm{s}}} + B \rvu_{k_{\mathrm{s}}} = \rvx_{k_{\mathrm{s}} +1} $ to derive the right hand side of this equation.
By applying the vec operator to \eqref{Bellman2.5}, we obtain
\begin{align}
&((\rvx_{k_{\mathrm{s}}}^\top \otimes \rvx_{k_{\mathrm{s}}}^\top) -(\rvx_{k_{\mathrm{s}}+1}^\top  \otimes \rvx_{k_{\mathrm{s}}+1}^\top) )\textrm{vec}(G^{0}_1) \notag \\
&+2(\rvx_{k_{\mathrm{s}}}^\top \otimes (\rvu_k-K^{\mathrm{init}} \rvx_{k_{\mathrm{s}}} )^\top )\textrm{vec} (G^{0}_2) \notag \\
&+((\rvu_k+K^{\mathrm{init}} \rvx_{k_{\mathrm{s}}})^\top \otimes (\rvu_k-K^{\mathrm{init}} \rvx_{k_{\mathrm{s}}} )^\top) \textrm{vec} (G^{0}_3)\notag \\
&=\rvx_{k_{\mathrm{s}}}^\top ( Q + (K^{\mathrm{init}})^\top R K^{\mathrm{init}} ) \rvx_{k_{\mathrm{s}}},
\label{Bellman3}
\end{align}
where we used the relation $\vv_{1}^{\top} M \vv_{2} = (\vv_{2}^\top \otimes \vv_{1}^{\top} )\textrm{vec} (M)$ derived from the properties of vec operator~\cite[\S 4.3.1]{hj1994topics}. 
In addition, $G^{0}_1 \in \sR^{n \times n}$, $G^{0}_2 \in \sR^{m \times n}$ and $G^{0}_3 \in \sR^{m \times m}$ are the matrices defined as follows:
\begin{align}
&G^{0}_1 : =P^{0} \label{L1}, \\ 
&G^{0}_2 := B^\top P^{0} A, \label{L2} \\
&G^{0}_3 := B^\top P^{0} B. \label{L3}
\end{align}
The same formula as \eqref{Bellman3} holds for the data collected at other time steps. By putting these equations together, we have the following system of $l$ linear equations with $n^2 + nm + m^2 $ unknowns: 
\begin{align}
  F^{0}[\textrm{vec}(G^{0}_1)^\top ,\textrm{vec}(G^{0}_2)^\top ,\textrm{vec}(G^{0}_3)^\top]^\top  
 =  \bm{h}^0. \label{linearequation}
\end{align}
When $l$ and $\bm{\nu}_{k}$ are appropriately chosen and the rank of $F^{0}$ is  $n^2 + nm + m^2 $, the exact and unique solution 
\begin{align}
  [\textrm{vec}(G^{0}_1)^\top ,\textrm{vec}(G^{0}_2)^\top ,\textrm{vec}(G^{0}_3)^\top]^\top  
 =  ((F^{0})^{\top} F^{0} )^{-1}(F^{0})^{\top}\bm{h}^0
\end{align}
is obtained.  
On the other hand, the Lyapunov equation 
\begin{align}
    &P^{0}-(A+BK^{\mathrm{init}})^\top P^{0} (A+BK^{\mathrm{init}})
    = Q + (K^{\mathrm{init}})^\top R K^{\mathrm{init}}
    \label{Ricatti1}
\end{align}
is obtained by substituting \eqref{StateEquationFB} into the Bellman equation~\eqref{Bellman}. 
Furthermore, by substituting \eqref{L2} and \eqref{L3} into \eqref{al1_K} in  Algorithm~\ref{alg1}, we obtain 
\begin{align}
     K^{1} = - ( B^\top P^{0} B + R )^{-1} B^\top P^{0} A.
     \label{Ricatti2}
\end{align}

As shown in~\cite{Difference_Riccati_equation_convergence}, $A + BK^{1}$ is Schur stable with this $K^{1}$. 
Therefore, the procedure described above can be continued recursively for $j = 1, 2, \ldots$ 
This also implies that updating $K^{j}$ with  \eqref{algorithm_AC} and \eqref{al1_K} is equivalent to updating it with \eqref{Ricatti3} and \eqref{Ricatti4} below:
\begin{align}
& P^{j}-(A+BK^{j})^\top P^{j} (A+BK^{j}) = Q + (K^{j})^\top R K^{j}, \label{Ricatti3} \\
& K^{j+1} = - ( B^\top P^{j} B + R )^{-1} B^\top P^{j} A, \label{Ricatti4}
\end{align}
which comprise the Hewer's iterative method for discrete-time algebraic Riccati equation~\cite{Difference_Riccati_equation_convergence}. 
Thus, the theoretical assurance regarding the updates by \eqref{Ricatti3} and \eqref{Ricatti4} given as Theorem 1 in  \cite{Difference_Riccati_equation_convergence}  also assures that $K^{j}$ updated by Algorithm~\ref{alg1} with the initial value $K^{\mathrm{init}}$ converges to the LQR solution $K^\star$ as $j \rightarrow \infty$.

In addition, the rate of convergence is quadratic in the neighborhood of the LQR solution $K^\star$ due to Theorem 2 in \cite{Difference_Riccati_equation_convergence}. 
\end{proof}

From Theorem~\ref{th:1}, an approximation of the optimal feedback gain near the equilibrium point of the nonlinear plant can be obtained by using Algorithm~\ref{alg1}. 

We show a specific example of $K^\mathrm{AC}$ designed by Algorithm~\ref{alg1} in Section~\ref{sec:example}.

\begin{remark}
Algorithm~\ref{alg1} is the discrete-time correspondent of the offline RL method for the continuous-time LQR problem proposed in \cite{jiang2012computational}, and also a special case of the offline RL method for discrete-time $H_{\infty}$ control problem proposed in \cite{Risan_step1} with simpler formulae.
However, Algorithm~\ref{alg1} and the proof of Theorem~\ref{th:1} are derived specifically and directly for the offline reinforcement learning of the discrete-time LQR solution based on the Hewer's method~\cite{Difference_Riccati_equation_convergence}. 
This leads to the algorithm suitable for the Step 1 of the model-free two-step design approach in the problem setting of this study, and also to revealing the rate of convergence of the algorithm that is not shown in the previous work~\cite{jiang2012computational,Risan_step1}.
\end{remark}

\begin{remark} \label{remark:prb_noise}
    It is difficult to explicitly state what kind of exploration term (probing noise) should be applied to make the rank condition regarding $F^j$ hold. 
    In fact, the related work~\cite{jiang2012computational,Risan_step1} did not provide such information in each problem setting either. Intuitively, the exploration term should contain a sufficiently large number of frequency components to avoid the degeneration of the matrix $F^j$. In the simulation in Section \ref{sec:robustness}, this is empirically confirmed using an exploration term consisting of various sine functions. 
\end{remark}

\begin{remark}
The RL method for the LQR problem was previously studied for both continuous-time and discrete-time systems. 
For continuous-time systems, Jiang and Jiang \cite{jiang2012computational} and Bian and Jiang \cite{Bian2016Value} proposed offline policy-iteration-based and value-iteration-based algorithms, respectively. Later, these RL algorithms were extended to design the gain for the output feedback case \cite{Rizvi2020Reinforcement}. 
For discrete time systems, online policy-iteration-based algorithms were proposed in literatures \cite{bradtke1994adaptive, Lewis2009Reinforcement, Lewis2011Reinforcement, Kiumarsi2014Linear}. Offline algorithm were proposed in literature \cite{fazel2018global,Donghwan2019Prime}, where the discrete-time LQR controller was obtained from multiple experimental data with different initial values. 
Compared to these related methods, the {\it offline} RL method for {\it discrete-time} LQR problem given as  Algorithm \ref{alg1} in this section is particularly suited for the redesign problem for two reasons: (i) Algorithm \ref{alg1} requires only a single experimental data for learning, minimizing the wear on the plant during the learning process, and (ii) the discrete-time cost function (\ref{costfunc}) used for the offline RL is compatible with that for the online RL used in Step 2 as opposed to continuous-time cost functions. 
\end{remark}

\section{DEMONSTRATION EXAMPLE}\label{sec:example}
In this section, we illustrate the validity of the proposed approach through numerical simulations using an inverted pendulum with input saturation as a plant. 

\subsection{Plant and overall flow of the simulation}
\begin{figure}
 \centering
 \includegraphics[width=12cm]{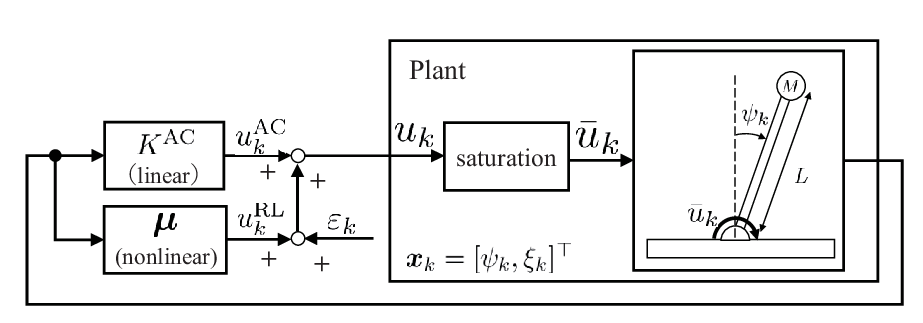}
 \caption{Plant used in the numerical simulation (inverted pendulum with input saturation). We design a nonlinear optimal regulator with the proposed model-free two-step design approach based on RL to control this inverted pendulum.}
 \label{fig:pendulum_block_diagram}
\end{figure}
We consider the control problem of the inverted pendulum shown in \rfig{fig:pendulum_block_diagram}.
This inverted pendulum is controlled by the torque input $\bar{u}_{k} \in \mathbb{R}$ from the motor attached to the fulcrum.
The motor has input saturation characteristics, by which the control input $u_{k}$ is saturated at a constant value of $\pm s$. 
That is, the actual torque input $\bar{u}_{k}$ applied to the fulcrum is
\begin{align}
\bar{u}_{k}= 
\left \{
\begin{array}{cl}
-s & {\rm if } \quad u_{k} < -s \\
u_{k} & {\rm if } \quad -s \le u_{k} \le s \\
s & {\rm if} \quad s < u_{k}
\end{array}
\right..
\label{saturation}
\end{align}

The discrete-time model of the dynamics of the inverted pendulum is given by
\begin{align}
\bm{x}_{k+1}=
\begin{bmatrix}
\psi_{k}  + T_{\mathrm{s}} \xi_{k}\\
\xi_{k} + \displaystyle\frac{g T_{\mathrm{s}}}{L}\sin \psi_k - \frac{\eta T_{\mathrm{s}}}{ML^2}\xi _k +\frac{T_{\mathrm{s}}}{ML^2}\bar{u}_{k}
\end{bmatrix}, 
\label{pendulum_system}
\end{align}
 where $T_{\mathrm{s}}$ is the sampling period, and $\bm{x} = [\psi, \xi]^\top$ is the state of the inverted pendulum consisting of  the angle (rad) and angular velocity of the pendulum (rad/s) denoted by $\psi \in [-\pi, \pi]$ and $\xi \in (-\infty, \infty)$, respectively. The definitions and the values of  other variables in \eqref{pendulum_system} are shown in Table \ref{tab:step1_parameters}.

Suppose the inverted pendulum is stabilized by a control law  $K^{\mathrm{init}}$ (Assumption \ref{supp1}), but the control performance is not satisfactory.
In what follows, our goal is to redesign the control law by applying the two-step procedure proposed in Section \ref{sec:controlaw}.
Specifically, we first regard the nonlinear system \eqref{pendulum_system} as a plant and obtain the quasi-optimal auxiliary control law $K^\mathrm{AC}$ through Algorithm \ref{alg1}, assuming that the model of the plant is unknown (Step 1). 
Then, with $K^\mathrm{AC}$, we design the nonlinear control law $\bm{\mu}$ by applying a general online RL method (Step 2). 
In this example, we use an Actor-Critic method with eligibility traces and describe it in detail in Section~\ref{step2-eg-sec} and Appendix~\ref{append2}.

We define the finite interval cumulative cost from time $k=0$ to $k=k_{\mathrm{fin}}$ as 
\begin{align}
    J_{\mathrm{fin}}(k_{\mathrm{fin}}) = \sum^{k_{\mathrm{fin}}}_{k=0} (\vx_{k}^\top Q \vx_{k} + R u_{k}^2)
    \label{cost_J_fin}
\end{align}
for the comparison of control performance.
Note that if the state and input have sufficiently converged at $k=k_{\mathrm{fin}}$, $\bm{x}_k \simeq [0,0]^\top$ and $u_{k} \simeq 0$ hold after the terminal time step $k_{\mathrm{fin}}$. In such case,  we have
\begin{align}
    \sum^{\infty}_{k=k_{\mathrm{fin}}+1} (\vx_{k}^\top Q \vx_{k} + R u_{k}^2)
    \simeq 0
\end{align}
and $J_{\mathrm{fin}}(k_{\mathrm{fin}}) \simeq J$. 
In other words, the cost $J_{\mathrm{fin}}(k_{\mathrm{fin}})$ is a measure almost equivalent to the cost function $J$.

\subsection{Step 1: design of auxiliary control law ${\it K}^{\mathrm{AC}}$}
\subsubsection{Evaluation conditions}
\begin{table}
 \caption{Simulation parameters of Step~1.}
 \label{tab:step1_parameters}
 \setlength{\tabcolsep}{3pt}
 \centering
  \begin{tabular}{p{60pt}p{225pt}p{60pt}}
   \hline 
   Symbol & Definition & Value \\
   \hline \hline
   $L$ & Length of the pendulum $(\mathrm{m})$  & $0.5$\\
   $M$ & Mass of the pendulum head $(\mathrm{kg})$ & $0.15$\\
   $g$ &  Gravitational constant  $(\mathrm{m/s^2})$ & $9.8$\\
   $\eta$ &  Friction coefficient & $0.05$\\
   $Q$ & Weighting matrix for state & $\mathrm{diag}[100, 1]$\\
   $R$ & Weighting matrix for input & $10$\\
   $K^{\mathrm{init}}$ & Initial control law & $[-8.23, -1.00]$\\   
   $T_{\mathrm{s}}$ & Sampling period $(\mathrm{s})$ & $0.06$\\
   $k_{\mathrm{fin}}$ & Terminal time step for the finite interval cumulative cost & 50 \\
   $l$ & Number of data  & $30$\\
   $\epsilon$ & Threshold for termination & $  1 \times 10^{-3}$\\
   $s$ & Saturation value of the torque $(\mathrm{N \cdot m})$ & $0.5$\\
   $a$ &  Amplitude parameter of the exploration term  & $0.01$\\
   \hline
  \end{tabular}
\end{table}
\normalsize

In Step 1, we run the iterations in Algorithm \ref{alg1} with the parameters listed in Table \ref{tab:step1_parameters}.
The initial  state for data collection is set to $\bm{x}_{0} = \bm{0}$, and the following exploration term $\nu_{k}$ is added to $K^{\mathrm{init}} \bm{x}_{k}$:
\begin{align}
    \nu_k = a \sum^{100}_{i=1} \sin (\omega_{i} T_{\mathrm{s}} k), \label{nu}
\end{align}
where $\omega_i~(i = 1,~2,~\cdots,~100)$ are selected randomly from $[-500,~500]$.

\subsubsection{Evaluation results}
\begin{figure}
    \centering
    \includegraphics[width=10cm]{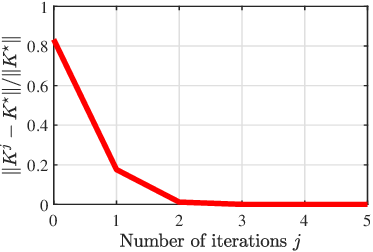}
    \caption{Convergence of $K^j$ updated by Algorithm~\ref{alg1}. $K^{j}$ converges to the linear optimal law $K^\star$. $K^{5}$ is selected as $K^\mathrm{AC}$ according to the termination condition \eqref{termination_condition} with $\epsilon$ given in Table~\ref{tab:step1_parameters}.}
    \label{fig:step1_convergence_K^j}
\end{figure}

The quasi-optimal linear control law $K^\mathrm{AC}$ is obtained as $[-2.77, -0.48]$.
For the purpose of verification, we also calculate the LQR controller $K^\star$ by using the linearized model of the nonlinear plant \eqref{pendulum_system} around the equilibrium point $[\psi, \xi]^{\top} = \bm{0}$. 
The result is $K^{\star}=[-2.77, -0.48]$.
As expected, $K^j$ converges to $K^\star$ by Algorithm \ref{alg1}, which is also  verified from \rfig{fig:step1_convergence_K^j} showing the gap between $K^j$ and $K^\star$ converges to $0$. 

Next, we compare the performance of the three control laws, $K^\mathrm{AC}$, $K^\star$, and $K^{\mathrm{init}}$, based on the average value of the cost $J_\mathrm{fin}(k_\mathrm{fin})$ obtained with 100 simulations. 
In these simulations, we select the initial state $\bm{x}_{0} = [\psi_{0}, \xi_{0}]^\top$ randomly according
to $\psi_{0} \in  [-0.4, 0.4]$ and $\xi_{0} \in [-1, 1]$. 
It can be seen in Table \ref{tab:Step1_cost} that the performance of the control law $K^\mathrm{AC}$ is better than that of the initially given control law $K^{\mathrm{init}}$. 
We can also see that the performance of $K^\mathrm{AC}$ matches that of $K^{\mathrm{\star}}$, which is a direct consequence of the convergence of $K^\mathrm{AC}$ to $K^{\mathrm{init}}$.  
\begin{table}
 \caption{Average of the cost $J_{\mathrm{fin}}(k_{\mathrm{fin}})$ for each linear control law. Average of the cost using $K^\mathrm{AC}$ is lower than using $K^\mathrm{init}$ and matches to the value using linear optimal controller $K^\star$ }
  \setlength{\tabcolsep}{3pt}
  \centering
  \begin{tabular}{p{100pt}p{120pt}p{200pt}}
  \hline
  Control law & Value & Average of the cost $J_{\mathrm{fin}}(k_{\mathrm{fin}})$\\
  \hline \hline
  $K^\mathrm{AC}$ & $ [-2.77, -0.48] $ & $ 39.0 $\\
  $K^{\mathrm{init}}$ & $[-8.23, -1.00] $ & $ 133.4 $\\
  $K^\star$ & $[-2.77, -0.48]$ & $ 39.0 $\\
  \hline
  \end{tabular}
   \label{tab:Step1_cost}
\end{table}
These results show the validity of Theorem~\ref{th:1}.

\subsection{Step 2: design of nonlinear control law $\bm{\mu}$}
\label{step2-eg-sec}
\subsubsection{Evaluation conditions}
In Step 2, the nonlinear control law $\bm{\mu}$, or more precisely the control law parameter $W$ is adjusted by an online RL method. 
In this demonstration example, we use an Actor-Critic method with eligibility traces combined with the linear control law \cite[\S 13.5]{sutton2018reinforcement}, \cite{okawa2019control} whose pseudo code is shown as Algorithm~\ref{alg2} in Appendix~\ref{append2}. 
To obtain a parameter value that can generate appropriate input for an arbitrary state, we perform the following training procedure. A training experiment simulation consists of  $N_{\mathrm{tri}} = 4000$ trials and each trial is performed for 
$k_{\mathrm{fin}}=50$ steps.
 We select the initial state $\vx_0 = [\psi_0, \xi_0]^\top$ randomly according to $\psi_0 \in [-0.4,0.4]$ and $\xi_0 \in [-1,1]$ in each trial. 
If the angle of the pendulum exceeds 0.5 rad, \textit{i.e.}, $|\psi_k| \geq 0.5$, 
the trial is terminated. In this case, $W_k$ and $\bm{\theta}_k$ are updated by Algorithm \ref{alg2} but by setting the reward to $r_k = -1000$ as a penalty and $\bm{\theta}_{k-1}^\top \bm{\phi}(\bm{x}_k) = 0$ (line 8 of Algorithm \ref{alg2}).

To make $W$ converge, we set  the variance  $\Sigma = \sigma^2$  of the exploration term in $j$-th trial to 
\begin{align}
 \sigma_{\rm init}^2 \times (10^{-4})^{\frac{j}{N_{\mathrm{tri}}}},
\label{eq:sigma}
\end{align}
and the learning rate $\beta$  for $W$ in $j$-th trial to
\begin{align}
     \beta_{\rm init} \times (10^{-2})^{\frac{j}{N_{\mathrm{tri}}}}
    \label{eq:_beta}
\end{align}
so that the degree of exploration and the change of $W$ decrease as the number of trials increases, where $\sigma_{\rm init}^2$ denotes the initial variance and $\beta_{\rm init}$ denotes the initial learning rate.

The change of the parameter $W$ and the cost $J_{\mathrm{fin}}(k_{\mathrm{fin}})$ is affected by the stochasticity of the exploration term and is also highly dependent on the sequence of initial states. 
Therefore, the trend of the transient control performance during learning needs to be evaluated in a statistical manner. 
For this purpose, we define a sequence of 4000 trials as one set of simulations and execute $N_\mathrm{sim}=3500$ sets of simulations to calculate the average of the cost  $J_{\mathrm{fin}}(k_{\mathrm{fin}})$ at each trial over the 3500 sets. 
In addition, we consider the following two cases for comparison: (i) the case where we train $\vmu$ by Algorithm~\ref{alg2} without any linear control law (hereafter denoted as RL alone) and (ii) the case where we use $K^{\rm init}$ instead of $K^{\rm AC}$ in Algorithm~\ref{alg2} (denoted as $K^{\rm init}$ + RL). 
The same procedure is performed for these two cases to compute the average of the transient learning performance.
We set the $i$-th basis function to 
\begin{align}
    \phi_i(\bm{x})= \exp  ( -\frac{||\bm{x}-\bm{c}_{i}||^2}{2\kappa_i^2} ),~~ i=1,2,\dots,N_{\mathrm{b}},  
\end{align}
where the average $\bm{c}_i$ is selected as  $\bm{c}_1 = [1,1]^\top, \bm{c}_2 =  [1,2]^\top, \ldots, \bm{c}_{120} =  [11,10]^\top, \bm{c}_{121} =  [11,11]^\top$.
The details of other simulation parameters are listed in Table \ref{step2_parameters}.

\begin{table}
 \caption{Simulation parameters of Step~2.}
 \label{step2_parameters}
 \setlength{\tabcolsep}{3pt}
 \centering
  \begin{tabular}{p{60pt}p{280pt}p{60pt}}
   \hline
   Symbol & Definition & Value \\
   \hline \hline
   $k_{\mathrm{fin}}$ &  Control time steps for one trial  & $50$\\
   $\kappa_i$ & Variance of $i$-th basis function & $0.5$ \\
   $\gamma$ & Discount rate  & $0.9$\\
   $\sigma_{\rm init}^2$ &  Initial variance of the exploration term & \\  
   ~ & ~~~~~~~~~~~~~~~~~~~~~~~~~~~~~ for $K^\mathrm{AC}$+RL  & $0.1$\\
    ~ & ~~~~~~~~~~~~~~~~~~~~~~~~~~~~~ for RL alone & $0.5$\\
   ~ & ~~~~~~~~~~~~~~~~~~~~~~~~~~~~~ for $K^{\mathrm{init}}$+RL & $5$\\    
   $\lambda ^\theta, \lambda ^\omega$ & Trace-decay rates & $0.99,0.99$\\
   $\alpha$ & Learning rate for state value weight & $0.05$\\
   $\beta_{\rm init}$ & Initial learning rate for nonlinear control  & \\ 
   ~ & law parameter  for $K^\mathrm{AC}$+RL & $0.0001$\\
   ~ & ~~~~~~~~~~~~~~~~ for RL alone & $0.0001$\\
   ~ & ~~~~~~~~~~~~~~~~ for $K^{\mathrm{init}}$+RL & $0.001$\\
   $N_{\mathrm{b}}$ & Number of basis functions  & $121$\\
   $N_{\mathrm{sim}}$ & Number of simulations  & $3500$\\
   $N_{\mathrm{tri}}$ & Number of trials comprising one simulation & $4000$\\
   \hline
  \end{tabular}
\end{table}

\begin{figure}
  \centering
  \includegraphics[width=12cm]{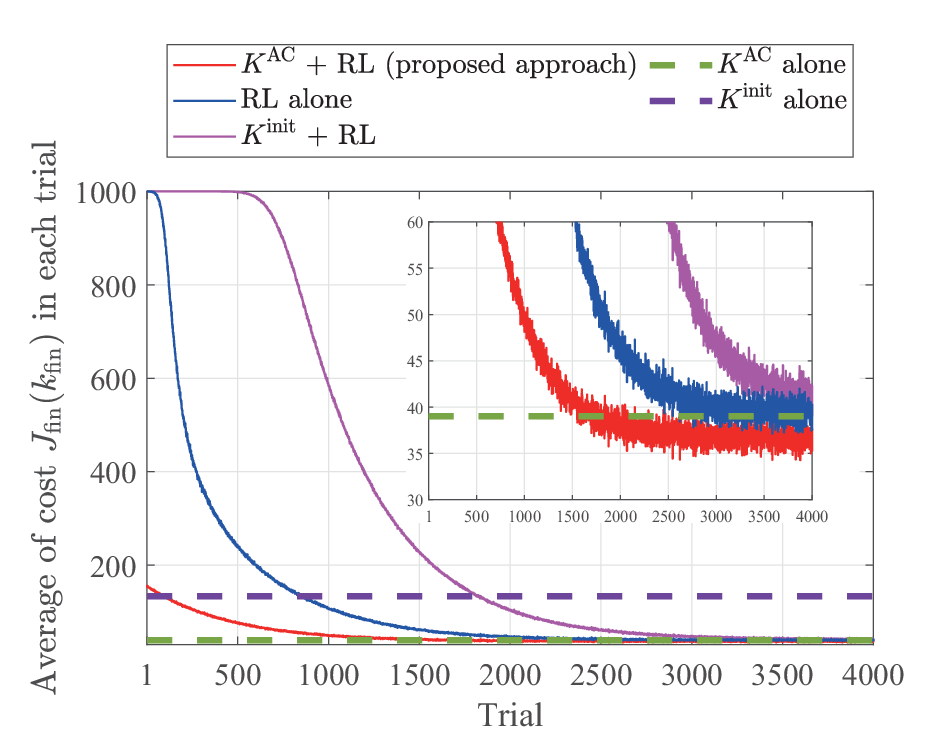}
    \caption{Averages of the cost $J_{\mathrm{fin}}(k_{\mathrm{fin}})$ in each trial in Step 2. An enlarged view with a different scale of the vertical axis is inserted to highlight the average cost at the end of the learning process. The proposed approach ($K^\mathrm{AC}$+RL) shows significantly better transient learning performance than that of RL alone especially in the early stage of learning. 
}
  \vspace{1mm}
  \label{fig:step2_Jfin}
\end{figure}

\subsubsection{Evaluation results}
The comparison of the average of the cost $J_{\mathrm{fin}}(k_{\mathrm{fin}})$ at each trial across 3500 sets of simulations is shown in \rfig{fig:step2_Jfin}, where the inset shows an enlarged view with a different range of the vertical axis. 
The cost for $K^\mathrm{AC}$ alone is calculated by averaging the cost for 3500 simulations with randomly selected initial states. 
It can be seen from \rfig{fig:step2_Jfin} that the average cost obtained with RL alone is very high in the early stage of learning. This is because the inverted pendulum falls over in a short time when the controller has little experience. 
On the other hand, the proposed approach ($K^\mathrm{AC}$+RL) shows significantly better transient learning performance. The average cost is lower than that of RL alone especially in the early stage of learning. 
This indicates that the quasi-optimal linear control law $K^\mathrm{AC}$ in the proposed approach assists the closed-loop system to maintain a certain level of  control performance  even in inexperienced states.
This interpretation agrees with the observation that the cost of using the online RL  in parallel with the initial control law ($K^{\mathrm{init}}$ +RL) is higher than that of the proposed approach in the early stage of learning since the control performance of the initial linear control law is worse than $K^\mathrm{AC}$. 
It should be noted that the cost $J_{\mathrm{fin}}(k_{\mathrm{fin}})$ for collecting the data in Step~1 is $4.35$, which is negligibly small compared with the cost for achieving the same control performance using RL alone and $K^{\mathrm{init}}$+ RL.
Therefore, designing $K^\mathrm{AC}$ in Step~1 is quite beneficial compared to the cost to obtain it online.   

Table \ref{tab:step2_last_cost} shows the average of the cost $J_{\mathrm{fin}}(k_{\mathrm{fin}})$ with the control laws designed by the proposed approach and four other comparison methods.
For each method, the average is computed with 3500 different control laws $\times$ 100 simulations starting from the same initial states as Table \ref{tab:Step1_cost}.
The nonlinear control law obtained by the proposed approach ($K^\mathrm{AC}$+RL) outperforms all the other control laws, \textit{i.e.,} $K^{\rm{init}}$ alone, $K^\mathrm{AC}$ alone, RL alone, and $K^\mathrm{init}$ + RL.
In particular, the combined quasi-optimal linear and nonlinear control law ($K^{\mathrm{AC}}$+RL) designed by the proposed approach improves the average cost $J_{\mathrm{fin}}(k_{\mathrm{fin}})$ by 7.2\% than the quasi-optimal linear control law $K^{\mathrm{AC}}$ alone, showing the effectiveness of the nonlinear controller $\bm{\mu}$ to attain better performance of the control system. 
Although the proposed approach requires additional costs for learning the nonlinear control law $\bm{\mu}$ compared to $K^{\mathrm{AC}}$ alone, this improvement of the average cost $J_{\mathrm{fin}}(k_{\mathrm{fin}})$ becomes dominant over the learning cost in the long run since the learning cost is incurred only once when the controller is trained. 

In conclusion, the proposed approach was shown to be effective for  improving transient learning performance and designing a control law with a smaller cost than RL alone under the same number of training trials.

\begin{table}
 \caption{Averages of the cost $J_{\mathrm{fin}}(k_{\mathrm{fin}})$ with control laws obtained by each method after Step 2. Average of the cost$J_{\mathrm{fin}}(k_{\mathrm{fin}})$ for the proposed approach is lower than other methods.}
 \centering
  \begin{tabular}{p{180pt}p{180pt}}
   \hline
   Method & Average of the cost $J_{\mathrm{fin}}(k_{\mathrm{fin}})$\\
   \hline \hline
   $K^\mathrm{AC}$ + RL (proposed approach) & $36.2$\\   
   $K^{\mathrm{init}}$ + RL & $40.3$ \\ 
   RL alone & $40.7$\\
   $K^\mathrm{AC}$ alone & $39.0$\\
   $K^{\rm{init}}$ alone & $133.4$\\
   \hline
  \end{tabular}
   \label{tab:step2_last_cost}
\end{table}

\begin{figure}
    \centering
        \includegraphics[width=8cm]{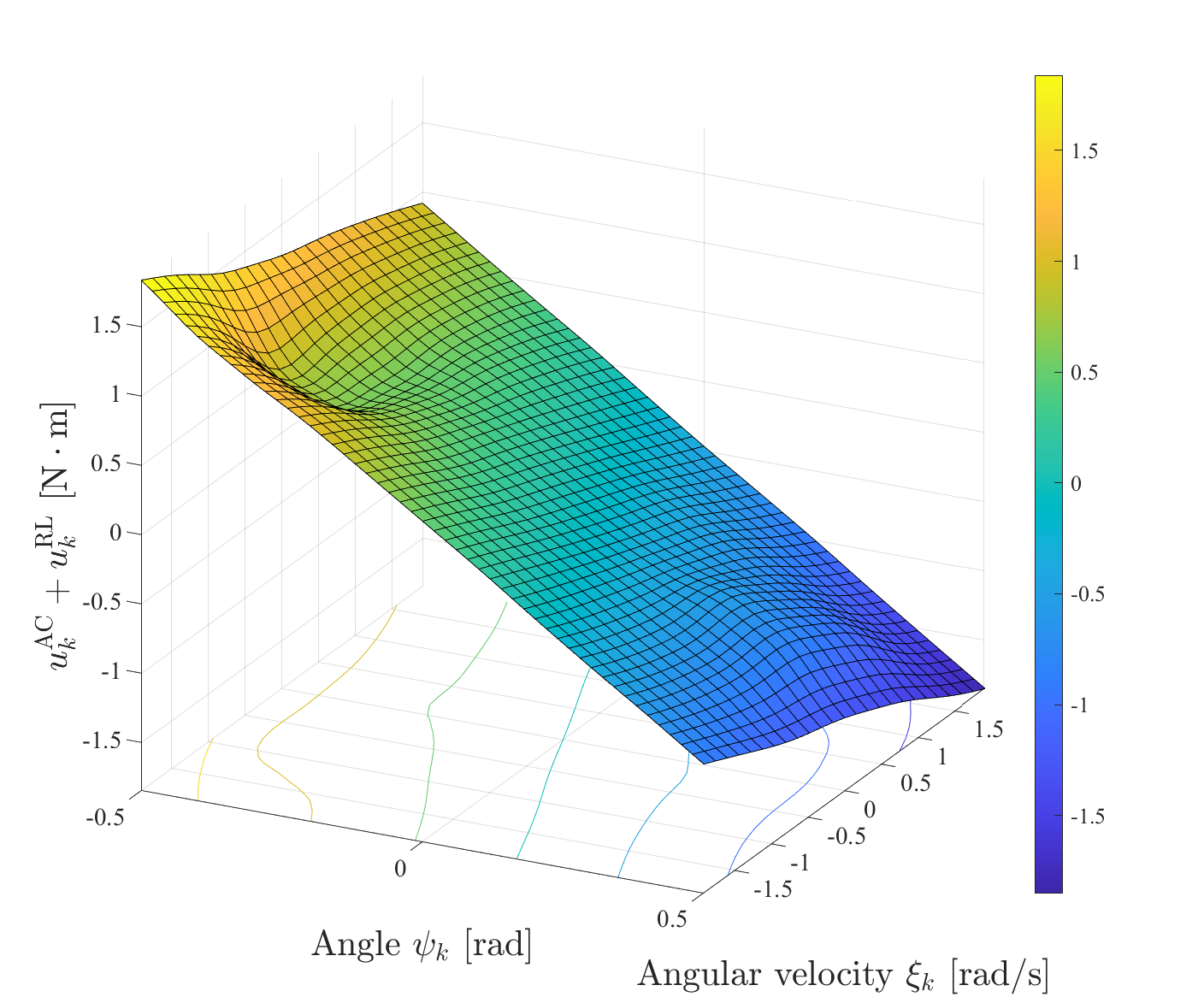}
    \caption{Visualization of the control law resulted from $K^\mathrm{AC}$ + RL (proposed approach). The designed control law is almost linear near the origin because the quasi-optimal liner control law  $K^\mathrm{AC}$ is dominant. On the other hand, the nonlinear control law $\bm{\mu}$ complements $K^\mathrm{AC}$ as the norm of the state tends to be large; it reduces the cost by suppressing the application of unnecessarily large control inputs that cause saturation.}
    \label{fig:policy_ACRL}
\end{figure}

\begin{figure}
\centering
\subfloat[Control input]{
\resizebox*{6cm}{!}{\includegraphics{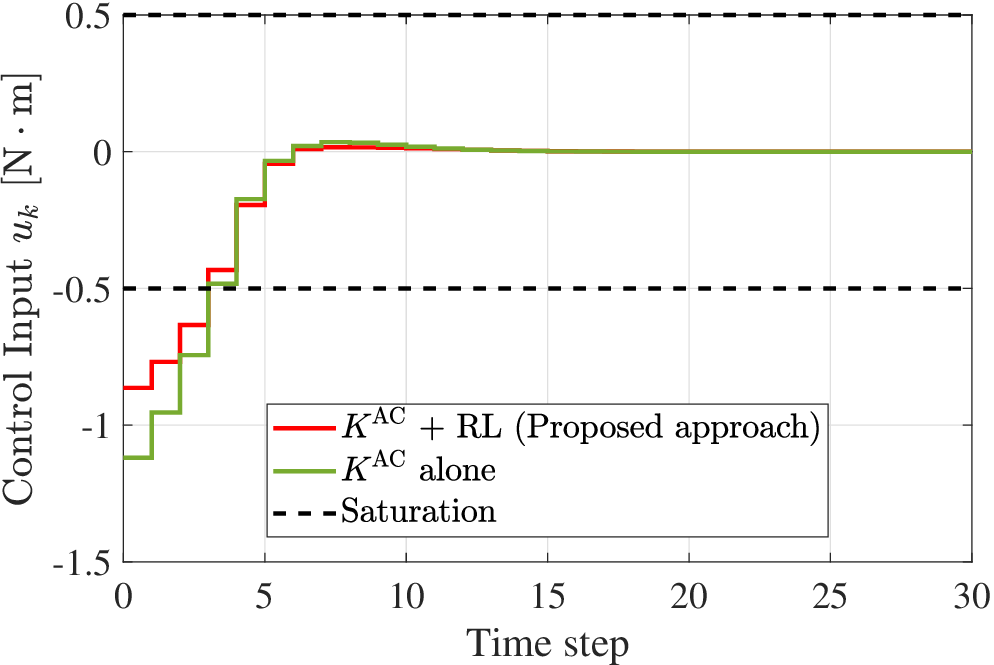}}}\hspace{5pt}
\subfloat[Actual torque input (saturated input)]{
\resizebox*{6cm}{!}{\includegraphics{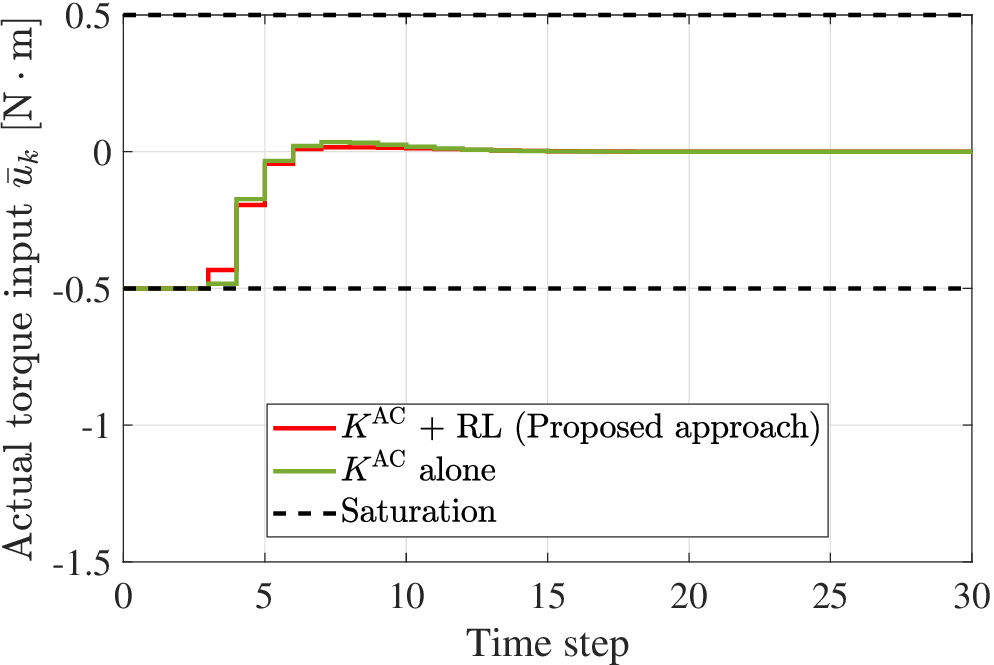}}}
\\
\subfloat[Angle]{
\resizebox*{6cm}{!}{\includegraphics{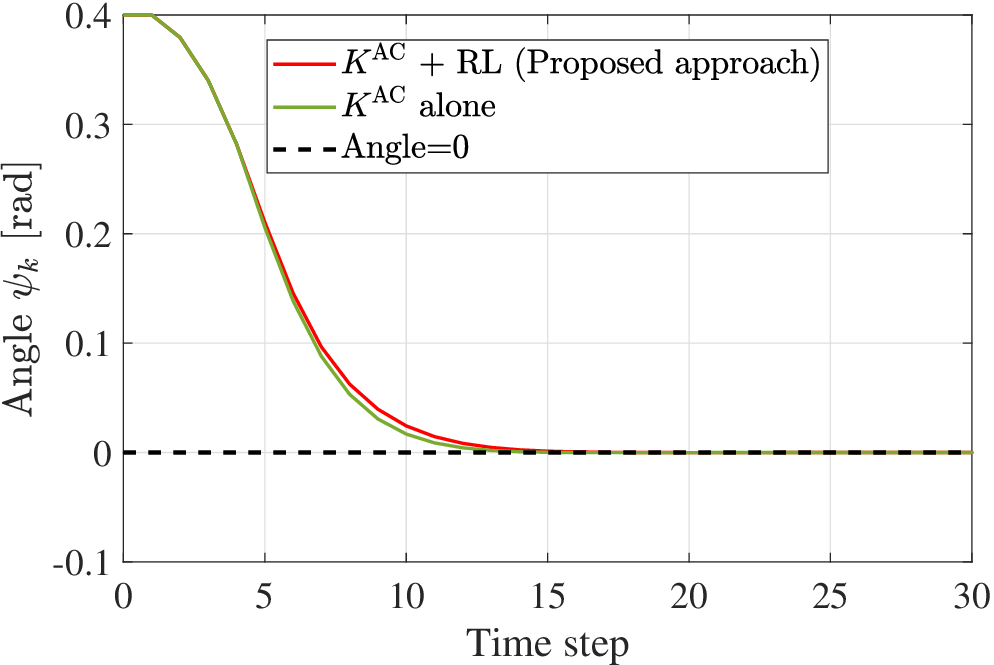}}}\hspace{5pt}
\subfloat[Angular velocity]{
\resizebox*{6cm}{!}{\includegraphics{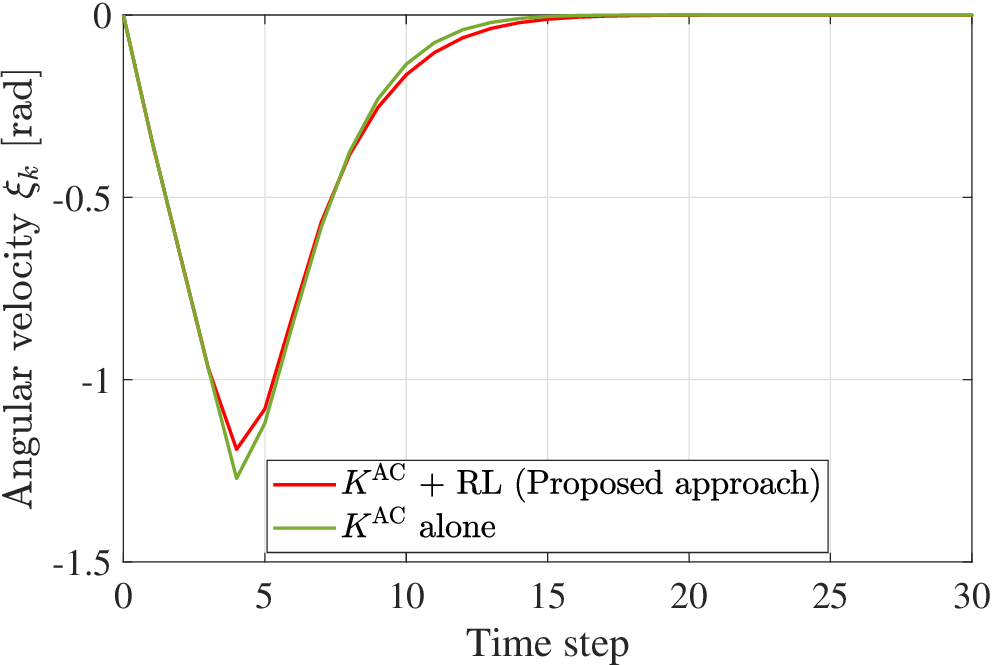}}}
\caption{The control input generated by the control law obtained by the proposed approach. The nonlinear control law $\bm{\mu}$ complements the liner control law $K^\mathrm{AC}$ as the norm of the state tends to be large; it reduces the cost by suppressing the application of unnecessarily large control inputs that cause saturation.} 
\label{fig:trajectory}
\end{figure}

To further interpret this result, we plot the control input generated by the control law obtained by the proposed approach in \rfig{fig:policy_ACRL}.
The figure illustrates that the designed control law is almost linear near the origin  because the nonlinear control law $\bm{\mu}$ is almost zero and  the quasi-optimal liner control law  $K^\mathrm{AC}$ is dominant. 
On the other hand, $\bm{\mu}$ complements $K^\mathrm{AC}$ as the norm of the state tends to be large; it reduces the cost by suppressing  the application of unnecessarily large control inputs that cause saturation (see Appendix~\ref{append1} for additional visualization).   
This can also be observed in the representative examples of the time-series of control input in \rfig{fig:trajectory}(a).
The corresponding state variables, \textit{i.e.,} the angle and the angular velocity of the pendulum, are shown in \rfig{fig:trajectory}(c),(d), where the experiment is executed with the initial  state $\bm{x}_{0} = [0.4,0]^\top$.
These figures show that the control law obtained by the proposed approach attains almost the same state trajectories as the quasi-optimal linear control law  $K^{\rm AC}$ but with the smaller control input by suppressing it for the saturated region.
Quantitatively, these differences sum up to the improvement of the average of the cost $J_{\mathrm{fin}}(k_{\mathrm{fin}})$ as shown in Table \ref{tab:step2_last_cost}.

\begin{remark}\label{remark:step2_add}
One of the major causes of wear on the plant is the operation of unstable system during the learning process in Step 2. Thus, to further reduce the wear during the learning process, {\it i.e}, to improve the transient learning performance, it is important to guarantee the stability of the closed-loop system during the learning process. 
Such an extension would be possible by modifying the structure of the nonlinear control law $\bm{\mu}$ as proposed in literature \cite{Samuele2021RL}. 
Specifically, we define the input to the plant $\bm{u}$ by 
\begin{align}
&    \bm{u}_k = K^{\mathrm{AC}} \bm{x}_k + {\bm{\mu}}(\bm{x}_k), \label{control-law-1} \\
&    {\bm{\mu}}(\bm{x}_k) := h(\bm{x}_k) (\hat{\bm{\mu}}(\bm{x}_k) - K^{\mathrm{AC}} \bm{x}_k),
\notag 
\end{align}
where $h$ and $\hat{\bm{\mu}}$ are functions to be learnt by any RL-based algorithms of user's choice in Step 2. 
Then, according to Theorem 1 in reference \cite{Samuele2021RL}, the control law (\ref{control-law-1}) makes the origin locally stable for any choice of the hyperparameters in $\hat{\bm{\mu}}$ under some assumptions (see \cite{Samuele2021RL} for details). 
This result can be regarded as an extension of the existing approach \cite{Samuele2021RL}  in that the model-based LQR controller in \cite{Samuele2021RL} can now be obtained from a single experimental data without knowing the information of the plant model using Algorithm \ref{alg1} of this paper. 
\end{remark}

\section{EFFICIENCY OF HYPERPARAMETER TUNING} \label{sec:robustness}
In the learning process of the RL-based control law, control inputs are generated in a stochastic manner, and thus, learning is not always successful; in the case of the inverted pendulum, for example, the resulting control law may fail to stabilize the pendulum, or it may deteriorate the performance compared to the initially given linear control law $K^{\mathrm{init}}$ even if the pendulum is stabilized. The ratio of successful learning depends on the setting of hyperparameters. In particular, the initial variance of the exploration term $\sigma_{\rm init}^2$ and the initial learning rate $\beta_{\rm init}$  are two major factors that directly affect the result. In this section, we show, through numerical simulations, that the stability and the performance of the control law obtained with the proposed approach ($K^\mathrm{AC}$+RL) are more robust against hyperparameter settings when compared to the control laws obtained  with the online RL method alone (RL alone) and the online RL method combined with the initially given linear control law  ($K^{\mathrm{init}}$+RL).

\subsection{Evaluation conditions}
We vary the initial variance of the exploration term $\sigma_{\rm init}^2$ in \eqref{eq:sigma} and the initial learning rate $\beta_{\rm init}$ in \eqref{eq:_beta}, and design $N_{\mathrm{sim}} = 100$ sets of control laws for each hyperparameter setting.
The percentage of (i) successful learning and (ii) improvement in performance is then evaluated for the three methods, $K^\mathrm{AC}$+RL (proposed), RL alone, and $K^{\mathrm{init}}$+RL. 
More specifically, we compute the average of the cost $J_{\mathrm{fin}}(k_{\mathrm{fin}})$ for 100 trials using each designed control law.
The percentage of successful learning is then calculated by $q/N_{\mathrm{sim}} \times 100$, where $q$ is the number of times satisfying the following two conditions:  (i) the parameter $W$ does not diverge to infinity at the end of the training process, and (ii) average of the cost is less than the pre-defined penalty cost for destabilization, which is set to 1000. 
The percentage of improvement in performance is calculated by $p/N_{\mathrm{sim}} \times 100$, where  $p$ is the number of times that average of the cost $J_{\mathrm{fin}}(k_{\mathrm{fin}})$ of using the trained control laws outperforms the quasi-optimal linear control law, $K^\mathrm{AC}$ alone.

\subsection{Evaluation results}
\begin{figure}[tb]
\centering
\subfloat[Percentage of successful learning in terms of stabilization]{
\resizebox*{6.5cm}{!}{\includegraphics{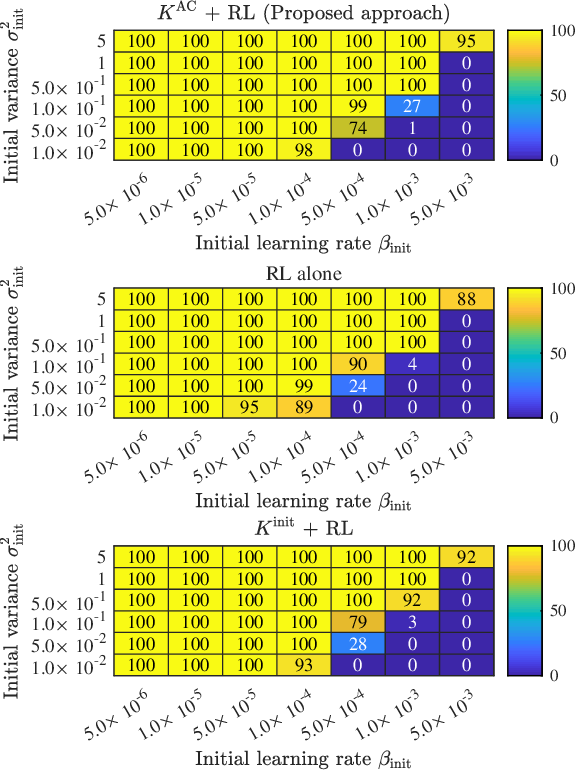}}}\hspace{5pt}
\subfloat[Percentage of improvement in performance over $K^{\rm AC}$ obtained by Algorithm~\ref{alg1}]{
\resizebox*{6.5cm}{!}{\includegraphics{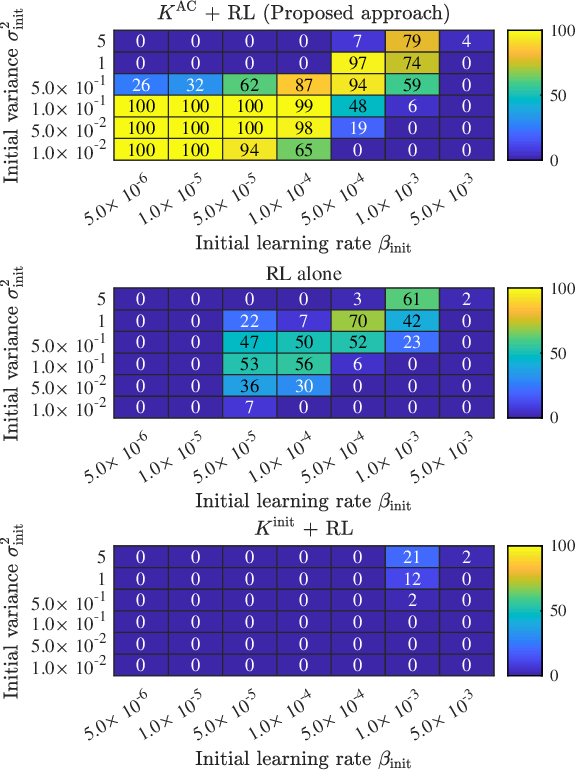}}}
\caption{
The percentage of (a) successful learning and (b) improvement in performance. In both figures, the results were obtained with (top) proposed approach, (middle) Online RL method alone, and (bottom) Online RL method using the initially given linear control law in parallel. The percentage of successful learning is almost the same between the three methods. On the other hand, the percentage of improvement for the proposed approach tends to be higher than other methods. The proposed approach in a wider range of hyperparameter combinations.
}
\label{fig:PoS_PoI}
\end{figure}

The percentages of (i) successful learning and (ii) improvement in performance are shown in Figs.~\ref{fig:PoS_PoI}(a) and \ref{fig:PoS_PoI}(b), respectively.
According to \rfig{fig:PoS_PoI}(a), the ratio of successful learning  is almost the same between the three methods, implying that dedicated tuning of the hyperparameters is not necessary for designing a stabilizing control law. 
On the other hand, \rfig{fig:PoS_PoI}(b) shows that hyperparameter tuning is important to obtain a better control law than the quasi-optimal linear one $K^{\rm AC}$. In this regard, the proposed approach is unmatched, \textit{i.e.,} the  control law obtained with the proposed approach shows better performance than $K^{\rm AC}$ for a wider range of hyperparameters compared with other methods.
This is because the control performance is almost optimized near the origin by $K^{\rm AC}$ designed in Step 1, and thus, the performance can be easily improved with a small initial variance $\sigma_{\rm init}^2$ and a learning rate $\beta_{\rm init}$, which determine the degree of exploration and the change in the control law parameter by the learning algorithm in Step 2. 
This feature allows us to avoid the tedious tuning of the hyperparameters and reduces wear on the plant in the controller design process.

\section{CONCLUSION}
\label{sec:conclusion}
In this paper, we have proposed a completely model-free approach to redesign the optimal regulator for nonlinear systems. Specifically, we have developed a model-free two-step design approach that improves the transient learning performance of RL and reduces the risk of wear on the plant during the learning process. 
To this goal, we have first developed an offline RL algorithm for designing a quasi-optimal linear control law. The quasi-optimal control law is then used to assist the control performance of the exploration phase of the online RL.  Using an inverted pendulum with input saturation as an example, we have shown that the proposed approach improves the transient learning performance of online RL and robustly achieves improvement in the performance of the control system for a wide range of hyperparameters.

\section*{Acknowledgements}
The authors would like to thank the anonymous reviewers for providing helpful feedback and literature information to improve our work.

\section*{Code Availability}
The official source code to reproduce the experimental results reported in Sections~\ref{sec:example} and \ref{sec:robustness} is publicly available at the following repository: \url{https://github.com/hori-group/two-step-design}. Please refer to the code for further details of the experiments.

\section*{Disclosure statement}
No potential conflict of interest was reported by the author(s).

\section*{Funding}
This work was supported by the JSPS KAKENHI under Grant JP18H01464. 

\section*{Notes on contributors}
Mei Minami received the B.E. degree in the Department of Applied Physico-informatics from Keio University in 2022.
Her research interests include control theory with RL.

Yuka Masumoto received the B.S. and M.S. degrees in engineering from Keio University in 2019 and 2021. In 2021, she began working as a Customer Success Account Manager at Microsoft Japan, taking a short break from research and working in a department with a mission of ``enabling digital transformation for organizations through assisting in customer success'' in the IT industry. Her research interests lie in feedback control theory and its fusion to machine learning/reinforcement learning.

Yoshihiro Okawa received the B.S., M.S., and Ph.D. degrees in engineering from Keio University, Tokyo, Japan, in 2012, 2013 and 2016, respectively. Currently, he works for Artificial Intelligence Laboratory, Fujitsu Limited. His research interests include robust control, distributed optimization, reinforcement learning and their applications. He is a recipient of Control Division Young Author’s Technology Award, Best Paper Award, and Control Division Young Author’s Award from SICE in 2020, 2017, and 2016, respectively, and is a Finalist of Young Author’s Award at SICE Annual Conference 2015. He is a member of SICE.

Tomotake Sasaki received his Bachelor’s degree in science from Waseda University, Japan, in 2004 and his Master’s and Ph.D. degrees in information science and technology from the University of Tokyo, Japan, in 2006 and 2010, respectively. After working at the University of Tokyo as a project academic support specialist, he joined Fujitsu Laboratories Ltd. in 2010. He was a visiting scientist at Massachusetts Institute of Technology from 2017 to 2018. He is a recipient of the 73rd IEEJ Academic Promotion Award Technical Development Award. He is currently a senior researcher at Fujitsu Limited, and a research affiliate at Massachusetts Institute of Technology and Center for Brains, Minds and Machines. His current research interests include control theory, reinforcement learning, deep learning and their applications. He is a member of IEEE, SICE, ISCIE, IEEJ, a committee member of IFAC TC 3.1 Computers for Control, and a NISTEP Expert in Science and Technology of National Institute of Science and Technology Policy.

Yutaka Hori received the B.S. degree in engineering, and the M.S. and Ph.D. degrees in information science and technology from the University of Tokyo in 2008, 2010 and 2013, respectively. He held a postdoctoral appointment at California Institute of Technology from 2013 to 2016. In 2016, he joined Keio University, where he is currently an associate professor. His research interests lie in feedback control theory and its applications to synthetic biomolecular systems. He is a recipient of Takeda Best Paper Award from SICE in 2015, and Best Paper Award at Asian Control Conference in 2011, and is a Finalist of Best Student Paper Award at IEEE Multi-Conference on Systems and Control in 2010. He has been serving as an associate editor of the Conference Editorial Board of IEEE Control Systems Society. He is a member of IEEE, SICE, and ISCIE.

\bibliographystyle{tfnlm} 
\bibliography{reference}

\begin{thebibliography}{10}
\providecommand{\url}[1]{\normalfont{#1}}
\providecommand{\urlprefix}{Available from: }

\bibitem{Hou2013}
Hou~Z, Wang~Z. From model-based control to data-driven control: Survey,
  classification and perspective. Information Sciences.
  2013;\hspace{0pt}235:3--35. {DOI}:
  \href{https://www.doi.org/10.1016/j.ins.2012.07.014}{10.1016/j.ins.2012.07.014}.

\bibitem{Hou2019}
Hou~Z, Xiong~S. On model-free adaptive control and its stability analysis. IEEE
  Transactions on Automatic Control. 2019;\hspace{0pt}64(11):4555--4569. {DOI}:
  \href{https://www.doi.org/10.1109/TAC.2019.2894586}{10.1109/TAC.2019.2894586}.

\bibitem{Hjalmarsson1998}
Hjalmarsson~H, Gevers~M, Gunnarsson~S, et~al. Iterative feedback tuning: theory
  and applications. IEEE Control Systems Magazine.
  1998;\hspace{0pt}18(4):26--41. {DOI}:
  \href{https://www.doi.org/10.1109/37.710876}{10.1109/37.710876}.

\bibitem{Campi2002}
Campi~M, Lecchini~A, Savaresi~S. Virtual reference feedback tuning: a direct
  method for the design of feedback controllers. Automatica.
  2002;\hspace{0pt}38(8):1337--1346. {DOI}:
  \href{https://www.doi.org/10.1016/S0005-1098(02)00032-8}{10.1016/S0005-1098(02)00032-8}.

\bibitem{Kaneko2013data}
Kaneko~O. Data-driven controller tuning: {FRIT} approach. In: Proc.~11th IFAC
  International Workshop on Adaptation and Learning in Control and Signal
  Processing; 2013. p. 326--336. {DOI}:
  \href{https://www.doi.org/10.3182/20130703-3-FR-4038.00122}{10.3182/20130703-3-FR-4038.00122}.

\bibitem{sutton2018reinforcement}
Sutton~RS, Barto~AG. Reinforcement learning: An introduction. 2nd ed. MIT
  Press; 2018.

\bibitem{Lewis2012_RL_and_FC}
Lewis~FL, Vrabie~D, Vamvoudakis~KG. Reinforcement learning and feedback
  control: Using natural decision methods to design optimal adaptive
  controllers. IEEE Control Systems Magazine. 2012;\hspace{0pt}32(6):76--105.
  {DOI}:
  \href{https://www.doi.org/10.1109/MCS.2012.2214134}{10.1109/MCS.2012.2214134}.

\bibitem{optimal_and_autonomous_control_using_RL}
Kiumarsi~B, Vamvoudakis~KG, Modares~H, et~al. Optimal and autonomous control
  using reinforcement learning: A survey. IEEE Transactions on Neural Networks
  and Learning Systems. 2018;\hspace{0pt}29(6):2042--2062. {DOI}:
  \href{https://www.doi.org/10.1109/TNNLS.2017.2773458}{10.1109/TNNLS.2017.2773458}.

\bibitem{bradtke1994adaptive}
Bradtke~SJ, Ydstie~BE, Barto~AG. Adaptive linear quadratic control using policy
  iteration. In: Proc. American Control Conference (ACC); 1994. p. 3475--3479.
  {DOI}:
  \href{https://www.doi.org/10.1109/ACC.1994.735224}{10.1109/ACC.1994.735224}.

\bibitem{doya2000reinforcement}
Doya~K. Reinforcement learning in continuous time and space. Neural
  Computation. 2000;\hspace{0pt}12(1):219--245. {DOI}:
  \href{https://www.doi.org/10.1162/089976600300015961}{10.1162/089976600300015961}.

\bibitem{murray2002adaptive}
Murray~JJ, Cox~CJ, Lendaris~GG, et~al. Adaptive dynamic programming. IEEE
  Transactions on Systems, Man, and Cybernetics, Part C: Applications and
  Reviews. 2002;\hspace{0pt}32(2):140--153. {DOI}:
  \href{https://www.doi.org/10.1109/TSMCC.2002.801727}{10.1109/TSMCC.2002.801727}.

\bibitem{Doya2002Multiple}
Doya~K, Samejima~K, Katagiri~Ki, et~al. Multiple model-based reinforcement
  learning. Neural Computation. 2002;\hspace{0pt}14(6):1347--1369. {DOI}:
  \href{https://doi.org/10.1162/089976602753712972}{10.1162/089976602753712972}.

\bibitem{Al-Tamimi2008}
Al-Tamimi~A, Lewis~FL, Abu-Khalaf~M. {Discrete-time nonlinear HJB solution
  using approximate dynamic programming: Convergence proof}. IEEE Transactions
  on Systems, Man, and Cybernetics, Part B: Cybernetics.
  2008;\hspace{0pt}38(4):943--949. {DOI}:
  \href{https://www.doi.org/10.1109/TSMCB.2008.926614}{10.1109/TSMCB.2008.926614}.

\bibitem{vamvoudakis2010online}
Vamvoudakis~KG, Lewis~FL. Online actor--critic algorithm to solve the
  continuous-time infinite horizon optimal control problem. Automatica.
  2010;\hspace{0pt}46(5):878--888. {DOI}:
  \href{https://www.doi.org/}{10.1109/TSMCC.2002.801727}.

\bibitem{sprangers2015reinforcement}
Sprangers~O, Babu^^c5^^a1ka~R, Nageshrao~SP, et~al. Reinforcement learning for
  port-{H}amiltonian systems. IEEE Transactions on Cybernetics.
  2015;\hspace{0pt}45(5):1017--1027. {DOI}:
  \href{https://www.doi.org/10.1109/TCYB.2014.2343194}{10.1109/TCYB.2014.2343194}.

\bibitem{Amherst2000}
Randl\o{}v~J, Barto~AG, Rosenstein~MT. Combining reinforcement learning with a
  local control algorithm. In: Proc. 17th International Conference on Machine
  Learning (ICML); 2000. p. 775^^e2^^80^^93782.

\bibitem{okawa2019control}
Okawa~Y, Sasaki~T, Iwane~H. Control approach combining reinforcement learning
  and model-based control. In: Proc.~12th Asian Control Conference; 2019. p.
  1419--1424. {URL}: \url{https://ieeexplore.ieee.org/document/8765068}.

\bibitem{Samuele2021RL}
Zoboli~S, Andrieu~V, Astolfi~D, et~al. Reinforcement learning policies with
  local lqr guarantees for nonlinear discrete-time systems. In: Proc. 60th IEEE
  Conference on Decision and Control (CDC); 2021. p. 2258--2263. {DOI}:
  \href{https://www.doi.org/10.1109/CDC45484.2021.9683721}{10.1109/CDC45484.2021.9683721}.

\bibitem{zanon2020safe}
Zanon~M, Gros~S. Safe reinforcement learning using robust {MPC}. IEEE
  Transactions on Automatic Control. 2021;\hspace{0pt}66(8):3638--3652. {DOI}:
  \href{https://www.doi.org/10.1109/TAC.2020.3024161}{10.1109/TAC.2020.3024161}.

\bibitem{xie2020model}
Xie~H, Xu~X, Li~Y, et~al. Model predictive control guided reinforcement
  learning control scheme. In: Proc.~2020 International Joint Conference on
  Neural Networks; 2020. {DOI}:
  \href{https://www.doi.org/10.1109/IJCNN48605.2020.9207398}{10.1109/IJCNN48605.2020.9207398}.

\bibitem{he2016deep}
He~K, Zhang~X, Ren~S, et~al. Deep residual learning for image recognition. In:
  Proc. 2016 IEEE/CVF Conference on Computer Vision and Pattern Recognition
  (CVPR); 2016. p. 770--778. {DOI}:
  \href{https://www.doi.org/10.1109/CVPR.2016.90}{10.1109/CVPR.2016.90}.

\bibitem{silver2018residual}
Silver~T, Allen~K, Tenenbaum~J, et~al. Residual policy learning [Arxiv
  preprint, arxiv:1812.06298]; 2018. {DOI}:
  \href{https://doi.org/10.48550/arXiv.1812.06298}{10.48550/arXiv.1812.06298}.

\bibitem{Difference_Riccati_equation_convergence}
Hewer~GA. An iterative technique for the computation of the steady state gains
  for the discrete optimal regulator. IEEE Transactions on Automatic Control.
  1971;\hspace{0pt}16(4):382--384. {DOI}:
  \href{https://www.doi.org/10.1109/TAC.1971.1099755}{10.1109/TAC.1971.1099755}.

\bibitem{hj1994topics}
Horn~RA, Johnson~CR. Topics in matrix analysis. 1st ed. Cambridge University
  Press; 1994.

\bibitem{jiang2012computational}
Jiang~Y, Jiang~ZP. Computational adaptive optimal control for continuous-time
  linear systems with completely unknown dynamics. Automatica.
  2012;\hspace{0pt}48(10):2699--2704. {DOI}:
  \href{https://www.doi.org/10.1016/j.automatica.2012.06.096}{10.1016/j.automatica.2012.06.096}.

\bibitem{Risan_step1}
Kiumarsi~B, Lewis~FL, Jiang~ZP. H${}_\infty$ control of linear discrete-time
  systems: Off-policy reinforcement learning. Automatica.
  2017;\hspace{0pt}78:144--152. {DOI}:
  \href{https://www.doi.org/10.1016/j.automatica.2016.12.009}{10.1016/j.automatica.2016.12.009}.

\bibitem{Bian2016Value}
Bian~T, Jiang~ZP. Value iteration and adaptive dynamic programming for
  data-driven adaptive optimal control design. Automatica.
  2016;\hspace{0pt}71:348--360. {DOI}:
  \href{https://doi.org/10.1016/j.automatica.2016.05.003}{10.1016/j.automatica.2016.05.003}.

\bibitem{Rizvi2020Reinforcement}
Rizvi~SAA, Lin~Z. Reinforcement learning-based linear quadratic regulation of
  continuous-time systems using dynamic output feedback. IEEE Transactions on
  Cybernetics. 2020;\hspace{0pt}50(11):4670--4679. {DOI}:
  \href{https://doi.org/10.1109/TCYB.2018.2886735}{10.1109/TCYB.2018.2886735}.

\bibitem{Lewis2009Reinforcement}
Lewis~FL, Vrabie~D. Reinforcement learning and adaptive dynamic programming for
  feedback control. IEEE Circuits and Systems Magazine.
  2009;\hspace{0pt}9(3):32--50. {DOI}:
  \href{https://doi.org/10.1109/MCAS.2009.933854}{10.1109/MCAS.2009.933854}.

\bibitem{Lewis2011Reinforcement}
Lewis~FL, Vamvoudakis~KG. Reinforcement learning for partially observable
  dynamic processes: Adaptive dynamic programming using measured output data.
  IEEE Transactions on Systems, Man, and Cybernetics, Part B (Cybernetics).
  2011;\hspace{0pt}41(1):14--25. {DOI}:
  \href{https://doi.org/10.1109/TSMCB.2010.2043839}{10.1109/TSMCB.2010.2043839}.

\bibitem{Kiumarsi2014Linear}
Kiumarsi~B, Lewis~FL, Modares~H, et~al. Reinforcement q-learning for optimal
  tracking control of linear discrete-time systems with unknown dynamics.
  Automatica. 2014;\hspace{0pt}50(4):1167--1175. {DOI}:
  \href{https://doi.org/10.1016/j.automatica.2014.02.015}{10.1016/j.automatica.2014.02.015}.

\bibitem{fazel2018global}
Fazel~M, Ge~R, Kakade~S, et~al. Global convergence of policy gradient methods
  for the linear quadratic regulator. In: Proc. 35th International Conference
  on Machine Learning; 2018. p. 1467--1476.
  \urlprefix\url{https://proceedings.mlr.press/v80/fazel18a.html}.

\bibitem{Donghwan2019Prime}
Lee~D, Hu~J. Primal-dual {Q}-learning framework for {LQR} design. IEEE
  Transactions on Automatic Control. 2019;\hspace{0pt}64(9):3756--3763. {DOI}:
  \href{https://doi.org/10.1109/TAC.2018.2884649}{10.1109/TAC.2018.2884649}.

\end{thebibliography}

\appendix
\section{Online RL for designing nonlinear control law}\label{append2}
The algorithm used for online RL in Step 2 in the numerical simulations is given in Algorithm \ref{alg2}, which is an Actor-Critic method with eligibility traces combined with the linear control law \cite[\S 13.5]{sutton2018reinforcement}, \cite{okawa2019control}.
\begin{algorithm}[t]
	\caption{Actor-Critic method with eligibility traces}
    \label{alg2}
	\textbf{Input}: $K^{\rm{AC}},\ \{ \bm{\phi}_{i} \}_{i=1}^{N_{\mathrm{b}}},\ \Sigma,\ \alpha,\ \beta,\ \gamma,\ \lambda^{\bm{\theta}},\  \lambda^{W},\ \bm{\theta}_0,\ W_0$ \\
	\textbf{Output}: $\bm{\theta},\ W$
	\begin{algorithmic}[1]
	\STATE Initialize eligibility traces as $ \bm{z}_{0}^{\bm{\theta}} = \bm{0}, Z_{0}^{W} = O, \zeta_{0} = 1$
	\STATE \textbf{while} a trial continues \textbf{do} the following at $k=0,1,\ldots$  
    \STATE \quad Observe state $\bm{x}_k$
     \STATE \quad \textbf{if} $k = 0$ \textbf{goto} line~\ref{GenerateAC}
     \STATE \quad \textbf{else} (\textit{i.e.,} $k \ge 1$)
     \STATE \quad \quad Get reward $r_{k}$
     \STATE \quad \quad Calculate TD error:  
     \STATE \quad \qquad $\delta_{k} = r_{k} + \gamma  \bm{\theta}_{k-1}^\top\bm{\phi}(\bm{x}_k)   - \bm{\theta}_{k-1}^\top \bm{\phi}(\bm{x}_{k-1})$
     \STATE \quad \quad Update eligibility traces: 
     \STATE \quad \qquad $\bm{z}_{k}^{\bm{\theta}} = \gamma \lambda^{\bm{\theta}} \bm{z}_{k-1}^{\bm{\theta}} + \zeta_{k-1} 
     \bm{\bm{\phi}}(\bm{x}_{k-1})$ 
     \STATE \quad \qquad {\small $Z_{k}^{W} = \gamma \lambda^{W} Z_{k-1}^{W} $} \\
        \quad \qquad \quad \quad {\small$ {} + \zeta_{k-1} \left. \frac{\partial \log{\rho}}{\partial W^\top} \right|_{\bm{u} = \bm{u}_{k-1},  \bm{x} = \bm{x}_{k-1}, W = W_{k-1}} $}
     \STATE \quad \qquad $\zeta_{k} = \gamma \zeta_{k-1}$
     \STATE \quad \quad Update state value weight: 
     \STATE \quad \qquad $\bm{\theta}_{k} = \bm{\theta}_{k-1} + \alpha \delta_{k} \bm{z}_{k}^{\bm{\theta}}$
     \STATE \quad \quad Update nonlinear control law parameter: 
     \STATE \quad \qquad $W_{k} = W_{k-1} + \beta \delta_{k} Z_{k}^{W}$
     \STATE \label{GenerateAC} \quad \quad Generate $\bm{u}^{\rm{AC}}_{k}$: $\bm{u}^{\rm{AC}}_{k} = K^{\rm{AC}} \bm{x}_{k}$
     \STATE \quad \quad Generate $\bm{u}^{\rm{RL}}_{k}$ according to Eq.~(\ref{u_RL})
     \STATE \quad \quad Generate $\bm{u}_{k}$: $\bm{u}_k = \bm{u}^{\rm{AC}}_{k} + \bm{u}^{\rm{RL}}_{k}$
     \STATE \quad \quad Input $\bm{u}_k$ to the plant
     \STATE \quad \textbf{end if}
     \STATE \textbf{end while}
     \STATE \textbf{if} a trial is terminated 
     \STATE \quad \quad \textbf{return} $\bm{\theta} = \bm{\theta}_{k},\ W = W_{k}$
     \STATE \textbf{end if}
     \end{algorithmic}
 \end{algorithm}

Specifically, we set the estimated value function $V$ for the state $\bm{x}$ to be 
\begin{align}
V(\bm{x};\bm{\theta}) = \bm{\theta}^\top \bm{\phi}(\bm{x}), \label{CRLMBC18}
\end{align}
where $\bm{\phi}(\bm{x}) := [\phi_1(\bm{x}), \cdots, \phi_{N_{\mathrm{b}}}(\bm{x})]^\top $ is the vector representation of basis functions $\{ \phi_{i} \}_{i=1}^{N_{\mathrm{b}}}$, $\bm{\theta} \in \mathbb{R}^{N_{\mathrm{b}}}$ is the state value weight, and $N_{\mathrm{b}}$ is the number of basis functions. We set the basis functions to
\begin{align}
    \phi_i(\bm{x})= \exp  ( -\frac{||\bm{x}-\bm{c}_{i}||^2}{2\kappa_i^2} ),~~ i=1,2,\dots,N_{\mathrm{b}},  
\end{align}
where $\bm{c}_i \in \mathbb{R}^n$ and $\kappa_{i}^2 > 0$ are the average and the variance of the $i$-th basis function, respectively.
The nonlinear control law generates inputs as follows:
\begin{align}
    \bm{u}^{\rm{RL}}_k = \bm{\mu} ( \vx_k; W_k ) + \bm{\varepsilon}_k.
\label{u_RL}
\end{align}
In this equation, we use a nonlinear control law 
$\bm{\mu}$ of the following form:
\begin{align}
    \bm{\mu} (\vx; W ) = W^\top \bm{\bm{\phi}}(\vx), 
\end{align}
which uses the same basis functions as $V$ and contains an adjustable parameter matrix $W\in \sR^{N_{\mathrm{b}} \times m} $.
Also, we generate the exploration term $\bm{\varepsilon}_k$ for online RL according to the Gaussian probability density function of the following form: 
\begin{align}
    \upsilon (\bm{\varepsilon}_k) 
    = \frac{1}{(2\pi)^{\frac{m}{2}} | \Sigma |^{\frac{1}{2}}} \exp ( -\frac{1}{2} \bm{\varepsilon}_k^\top \Sigma^{-1} \bm{\varepsilon}_k ). \label{CRLMBC20}
\end{align}
This means that $\bm{u}^{\rm{RL}}_k$ follows the Gaussian probability density function 
\begin{align}
    \lefteqn{\rho (\bm{u}^{\rm{RL}}_k | \vx_k ; W_k) = \frac{1}{(2\pi)^{\frac{m}{2}} | \Sigma |^{\frac{1}{2}}}} \nonumber \\
    & \times \exp ( -\frac{1}{2} (\bm{u}^{\rm{RL}}_k - \bm{\mu} (\vx_k; W_k ) )^\top \Sigma^{-1} (\bm{u}^{\rm{RL}}_k - \bm{\mu} (\vx_k; W_k )))
\label{policy_rho}
\end{align}
\noindent
with mean $\bm{\mu} (\vx_k; W_k )$ and covariance matrix $\Sigma$.
During the learning process, $\bm{u}_{k}^{\rm{RL}}$ is selected stochastically according to \eqref{u_RL}, and when the learning is finished, $\bm{u}_{k}^{\rm{RL}}=\bm{\mu} (\bm{x}_{k}; W)$ is used deterministically as the nonlinear control law with the learned parameter $W$. 
The parameters of both the state value function and the nonlinear control law are updated according to Algorithm \ref{alg2}, where $\bm{z}^{\bm{\theta}}_{k} \in \mathbb{R}^N$ and $Z^{W}_{k} \in \mathbb{R}^{N_{b} \times m}$ represent the eligibility traces for $\bm{\theta}$ and $W$, respectively, and $\lambda^{\bm{\theta}} \in [0,1)$ and $\lambda^{W} \in [0,1)$ are their trace-decay parameters. In addition, $\delta_{k} \in \mathbb{R}$ represents the temporal difference (TD) error, $\gamma \in [0,1]$ represents the discount rate, and $\alpha \in (0,1)$ and $\beta \in (0,1)$ represent learning rates. The instantaneous reward $r_k$ is defined by 
\begin{align}
r_k = -(\bm{x}_{k}^\top Q \bm{x}_{k} + \bm{u}_{k-1}^\top R \bm{u}_{k-1}),
\label{reward}
\end{align}
where $Q$ and $R$ are set to the same values as in Step 1 since this algorithm is designed to maximize a cumulative reward.
The inputs $\bm{\theta}_0\in\mathbb{R}^{N_{b}}$ and $W_0\in\mathbb{R}^{N_{b}\times m}$ are set arbitrarily for the first trial (episode); otherwise, they are set with the outputs $\bm{\theta}$ and $W$ obtained in the previous  trial.

\section{Visualization of the control law designed by the proposed approach}\label{append1}

\begin{figure}
\centering
\subfloat[Plot of $u^{\mathrm{AC}}$ generated by the linear control law $K^{\mathrm{AC}}$]{
\resizebox*{6.5cm}{!}{\includegraphics{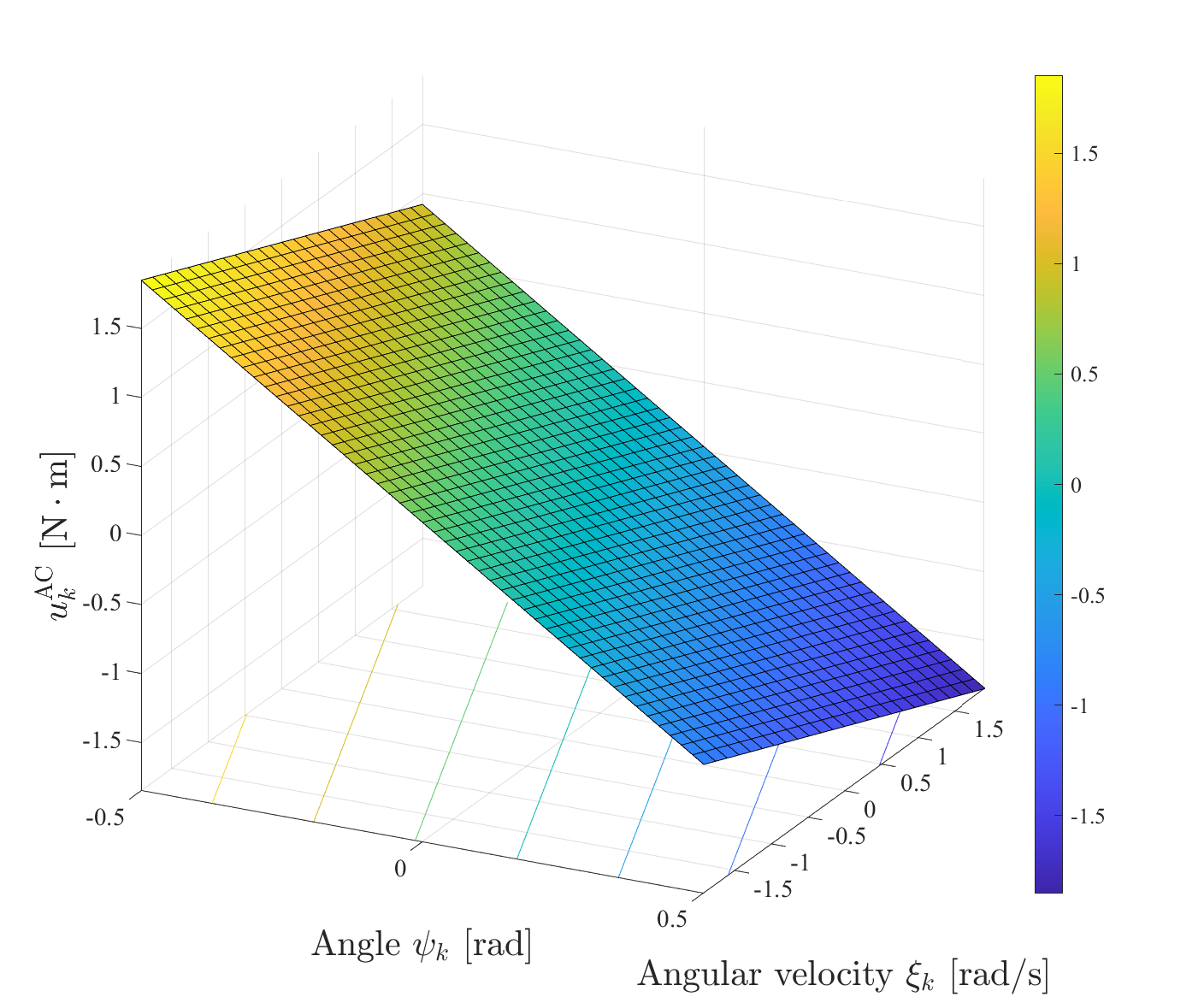}}}\hspace{5pt}
\subfloat[Plot of $u^{\mathrm{RL}}$ generated by the nonlinear control law $\bm{\mu}$]{
\resizebox*{6.5cm}{!}{\includegraphics{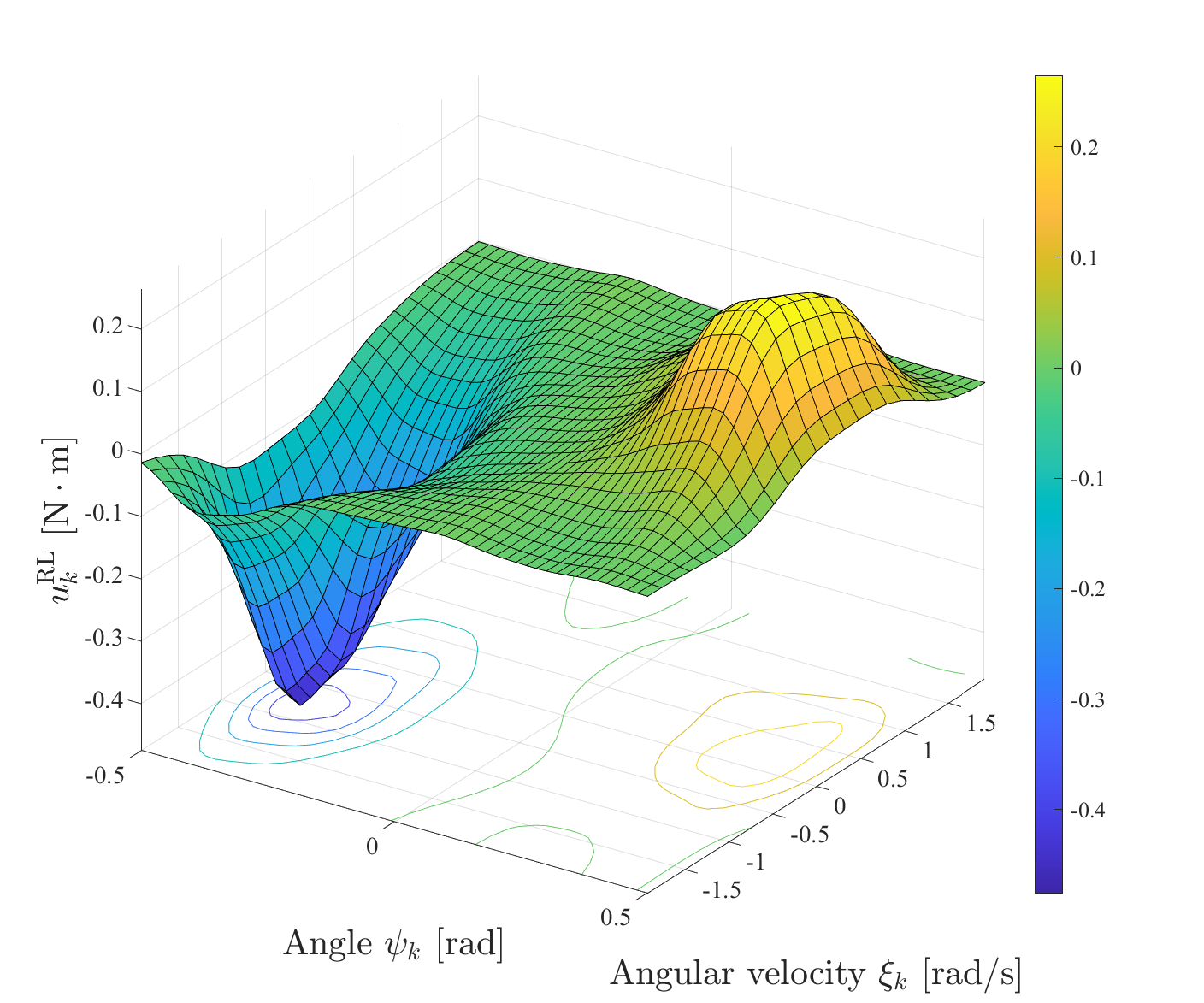}}}
\caption{Visualization of the (a) linear and (b) nonlinear part of the control law resulted from $K^{\mathrm{AC}}$ + RL (Proposed approach). The surface shown in (a) is flat. This is because  $K^{\mathrm{AC}}$ is a linear control law. 
On the other hand, the inputs in (b) (and \rfig{fig:policy_ACRL}) are generated by the nonlinear control law $\bm{\mu}$ in such a way that $ u^{\mathrm{AC}} + u^{\mathrm{RL}}$ becomes closer to $\pm s \ (\pm 0.5)$ when $u^{\mathrm{AC}}$ is too large / too small. 
In other words, the inputs $u^{\mathrm{RL}}$ are generated so that they are complementary to $u^{\mathrm{AC}}$ in terms of reducing the cost.} 
\label{fig:policy_appendix}
\end{figure}

In Subsection~\ref{step2-eg-sec}, \rfig{fig:policy_ACRL} shows the visualization of the inputs generated by the control law obtained by the proposed approach after Step~2. 
Here, \rfig{fig:policy_appendix}(a) and \rfig{fig:policy_appendix}(b) show the visualization of the inputs $u^{\mathrm{AC}}$ and $u^{\mathrm{RL}}$ generated by $K^{\mathrm{AC}}$ and $\bm{\mu}$, respectively.
We can see that the surface shown in  \rfig{fig:policy_appendix}(a) is flat. This is because  $K^{\mathrm{AC}}$ is a linear control law. 
On the other hand, we can see from \rfig{fig:policy_appendix}(b) (and \rfig{fig:policy_ACRL}) that  the inputs are generated by the nonlinear control law $\bm{\mu}$ in such a way that $ u^{\mathrm{AC}} + u^{\mathrm{RL}}$ becomes closer to $\pm s \ (\pm 0.5)$ when $u^{\mathrm{AC}}$ is too large / too small. 
In other words, the inputs $u^{\mathrm{RL}}$ are generated so that they are complementary to $u^{\mathrm{AC}}$ in terms of reducing the cost.

\end{document}